\documentclass[useAMS,usenatbib]{mn2e}

\usepackage{latexsym,graphicx,natbib}

\usepackage{color}
\usepackage[T1]{fontenc} 
\usepackage{aecompl}

\def\changed{}
\def\changed{\color{red}}

\newcommand\kms{{\rm\,km\,s^{-1}}}
\newcommand\msun{\rm\,M_\odot}

\newcommand\lsun{\rm\,L_\odot}
\newcommand\hii{H\,{\sc ii} \,}

\newcommand\myr{\msun \, {\rm yr}^{-1}}
\DeclareRobustCommand{\ion}[2]{%
\relax\ifmmode
\ifx\testbx\f
{\mathrm{#1\,\textsc{#2}}}\else {\mathrm{#1\,\mathsc{#2}}}\fi
\else\textup{#1\,{\mdseries\textsc{#2}}}%
\fi}
\newcommand{\MC}{\multicolumn}

\def\apgt{\ {\raise-.5ex\hbox{$\buildrel>\over\sim$}}\ }
\def\aplt{\ {\raise-.5ex\hbox{$\buildrel<\over\sim$}}\ }

\title[IRAS\,18153$-$1651]{IRAS\,18153$-$1651: an \hii region with a possible wind bubble blown by a young main-sequence B 
star\footnotemark[0]\thanks{Based on observations obtained at the Gemini Observatory (processed using the Gemini {\sc iraf} package), which is operated by the Association of Universities for Research in Astronomy, Inc., under a cooperative agreement with the NSF on behalf of the Gemini partnership: the National Science Foundation (United States), the National Research Council (Canada), CONICYT (Chile), Ministerio de Ciencia, Tecnolog\'{i}a e Innovaci\'{o}n Productiva (Argentina), and Minist\'{e}rio da Ci\^{e}ncia, Tecnologia e Inova\c{c}\~{a}o (Brazil),
and  observations collected at the Centro 
Astron\'omico Hispano Alem\'an (CAHA), operated jointly by the Max-Planck 
Institut f\"ur Astronomie and the Instituto de Astrofisica de Andalucia (CSIC).
} }
\author[V.V.Gvaramadze et al.]
       {V. V.~Gvaramadze,$^{1,2,3}$\thanks{E-mail: vgvaram@mx.iki.rssi.ru} 
       J.~Mackey,$^{4,5}$ A. Y.~Kniazev,$^{6,7,1}$ N.~Langer,$^{5}$ \\ 
       \newauthor 
       A.-N.~Chen\'e,$^{8} $ N.~Castro,$^{9}$ T. J.~Haworth$^{10}$ and E. K.~Grebel$^{11}$ \\
        $^{1}$Sternberg Astronomical Institute, Lomonosov Moscow State University, Universitetskij Pr. 13, Moscow 119992, Russia\\
        $^{2}$Space Research Institute, Russian Academy of Sciences, Profsoyuznaya 84/32, 117997 Moscow, Russia \\
        $^{3}$Isaac Newton Institute of Chile, Moscow Branch, Universitetskij Pr. 13, Moscow 119992, Russia \\
        $^{4}$Dublin Institute for Advanced Studies, Dunsink Observatory, Dunsink Lane, Castleknock, Dublin 15, Ireland \\
        $^{5}$Argelander-Institut f\"ur Astronomie, Auf dem H\"ugel 71, 53121 Bonn, Germany \\
        $^{6}$South African Astronomical Observatory, PO Box 9, 7935 Observatory, Cape Town, South Africa \\
        $^{7}$Southern African Large Telescope Foundation, PO Box 9, 7935 Observatory, Cape Town, South Africa \\
        $^{8}$Gemini Observatory, Northern Operations Center, 670 North A'ohoku Place, Hilo, HI 96720, USA \\
        $^{9}$Department of Astronomy, University of Michigan, 1085 S. University Avenue, Ann Arbor, MI 48109, USA \\
        $^{10}$Astrophysics Group, Imperial College London, Blackett Laboratory, Prince Consort Road, London SW7 2AZ, UK \\
        $^{11}$Astronomisches Rechen-Institut, Zentrum f\"ur Astronomie der Universit\"at Heidelberg, M\"onchhofstr. 
        12-14, D-69120 \\ 
        Heidelberg, Germany
        }
\begin{document}

\date{Accepted 2016 December 12. Received 2016 December 12; in original form 2016 October 24}


\maketitle

\label{firstpage}

\begin{abstract}
We report the results of spectroscopic observations and numerical modelling of the \hii region IRAS\,18153$-$1651. 
Our study was motivated by the discovery of an optical arc and two main-sequence stars of spectral type B1 and B3 
near the centre of IRAS\,18153$-$1651. We interpret the arc as the edge of the wind bubble (blown by the B1 star),
whose brightness is enhanced by the interaction with a photoevaporation flow from a nearby molecular cloud. This 
interpretation implies that we deal with a unique case of a young massive star (the most massive member of a 
recently formed low-mass star cluster) caught just tens of thousands of years after its stellar wind has begun to 
blow a bubble into the surrounding dense medium. Our two-dimensional, radiation-hydrodynamics simulations of the 
wind bubble and the \hii region around the B1 star provide a reasonable match to observations, both in terms of 
morphology and absolute brightness of the optical and mid-infrared emission, and verify the young age of 
IRAS\,18153$-$1651. Taken together our results strongly suggest that we have revealed the first 
example of a wind bubble blown by a main-sequence B star.
\end{abstract}


\section{Introduction}
\label{sec:intro}

Hot massive stars are sources of fast line-driven winds (Snow \& Morton 1976; Puls, Vink \& Najarro 2008), whose 
interaction with the circum- and interstellar  medium results in the origin of bubbles and shells of various shapes 
and scales (Johnson \& Hogg 1965; Lozinskaya \& Lomovskij 1982; Chu, Treffers \& Kwitter, 1983; Dopita et al. 1994). 
Since most (if not all) massive stars form in a clustered way (Lada \& Lada 2003; Gvaramadze et al. 2012), the wind 
bubbles produced by individual members of star clusters are unobservable because they merge into a single much more 
extended structure -- a superbubble (McCray \& Kafatos 1987). Numerous examples of such superbubbles (with diameters 
ranging from several tens of pc to kpc scales) were revealed in H$\alpha$ photographic surveys of the Magellanic 
Clouds and other nearby galaxies (e.g. Davies, Elliott \& Meaburn 1976; Meaburn 1980; Courtes et al. 1987; Hunter 1994).

To produce a well-shaped circular bubble or shell a massive star should be isolated from the destructive influence of 
winds from other massive stars (cf. Naz\'e et al. 2001). To achieve this, it should leave the parent cluster either 
because of a few-body dynamical encounter with other massive stars (e.g. Poveda, Ruiz \& Allen 1967; Oh \& Kroupa 2016) 
or binary supernova explosion (e.g. Blaauw 1961; Eldridge, Langer \& Tout 2011), or it should be the only massive star 
in the parent cluster. In the first case, the wind bubble around the star running away from its birth place rapidly 
becomes elongated (Weaver et al. 1977) and transforms into a bow shock (van Buren \& McCray 1988) if the star is moving 
supersonically with respect to the local interstellar medium (ISM). Stellar motion alone is hence the main reason that 
closed structures around hydrogen-burning massive field stars are not observed\footnote{Note that a runaway 
massive star can produce a short-living ($\sim10^4$ yr) circular shell during the advanced stages of evolution if it is 
surrounded by a dense material comoving with the star, i.e. the dense matter shed during the red supergiant phase, and if 
it is massive enough to evolve afterwards into a Wolf-Rayet star (Gvaramadze et al. 2009).}. The only known possible 
exception is the Bubble Nebula (e.g. Christopoulou et al. 1995; Moore et al. 2002), which is produced by the runaway 
O6.5(n)fp (Sota et al. 2011) star BD+60\,2522, whose luminosity class, however, is unknown because of the peculiar shape 
of the He\,{\sc ii} $\lambda$4686 line. The origin of this nebula might be attributed to a situation in which a 
bow-shock-producing star encounters a density enhancement (cloudlet) on its way, resulting in a temporal formation of a 
closed bubble with the star located near its leading edge. 

\begin{figure*}
\begin{center}
\includegraphics[width=14cm,angle=0]{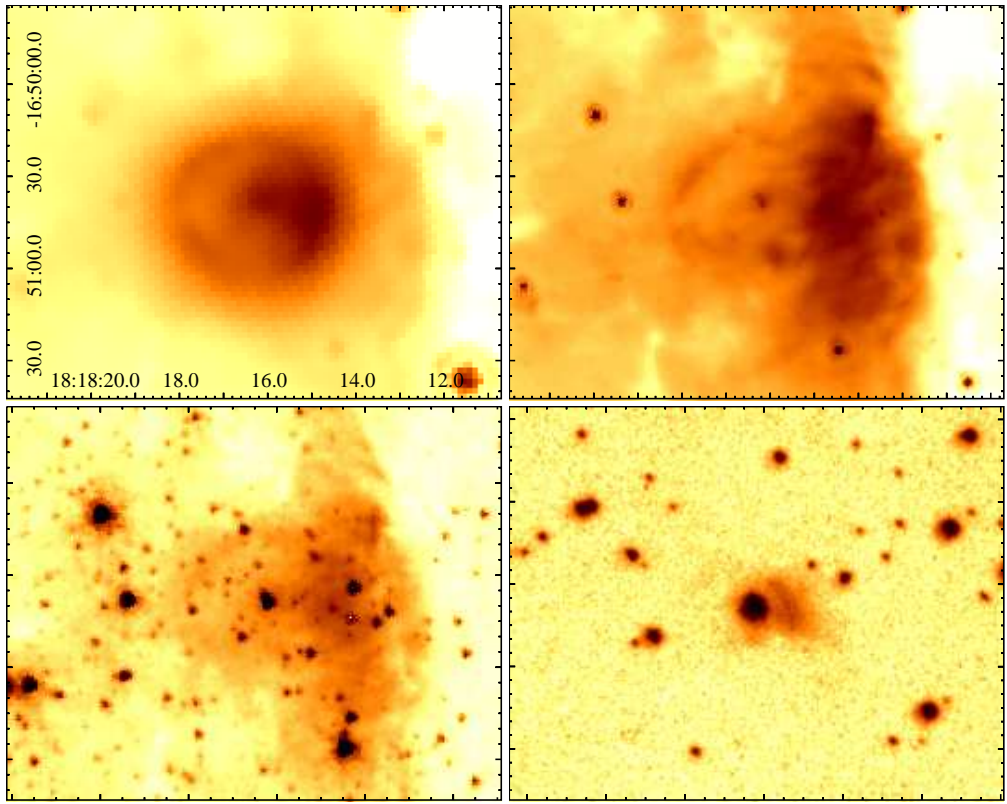}
\end{center}
\caption{From left to right, and from top to bottom: {\it Spitzer} MIPS $24 \, \mu$m 
and IRAC 8 and 3.6\,$\mu$m, and SHS H$\alpha$+[N\,{\sc ii}] images of IRAS\,18153$-$1651 
and its central stars (the scale and orientation of the images are the same). The 
coordinates are in units of RA (J2000) and Dec. (J2000) on the horizontal and vertical 
scales, respectively. At a distance of 2 kpc, 1 arcmin corresponds to $\approx$0.57 pc.
    }
\label{fig:neb}
\end{figure*}

In the second case, the wind-blowing star is the only massive star (a B star of mass of $8-10 \, \msun$) in a star cluster 
of mass of about $100 \, \msun$ (e.g. Kroupa et al. 2013). Such stars with their weak winds produce momentum-driven bubbles 
(Steigman, Strittmatter \& Williams 1975) in the dense material of the parental molecular cloud, which could be detected 
under favourable conditions, e.g. if the cluster was formed near the surface of the cloud. In this paper, we report the 
discovery of an optical arc within the circular mid-infrared shell (known as IRAS\,18153$-$1651) and argue that it represents 
the edge of a young ($\sim10^4$ yr) wind bubble produced by a main-sequence B star residing in a low-mass star cluster. 
In Section\,\ref{sec:bub}, we present the images of the arc, IRAS\,18153$-$1651 and two stars associated with them, as well 
as review the existing data on these objects. In Section\,\ref{sec:obs}, we describe our optical spectroscopic observations 
of the arc and the two stars. In Section\,\ref{sec:sta}, we classify the stars and model their spectra. 
In Section\,\ref{sec:arc}, we derive some parameters of the arc and propose a scenario for the origin of the arc and the shell 
around it. In Section\,\ref{sec:num}, we present and discuss results of numerical modelling, which we carried out to support 
the scenario. We summarize and conclude in Section\,\ref{sec:sum}.

\section{IRAS\,18153$-$1651 and its central stars}
\label{sec:bub}

\begin{figure*}
\begin{center}
\includegraphics[width=14cm,angle=0]{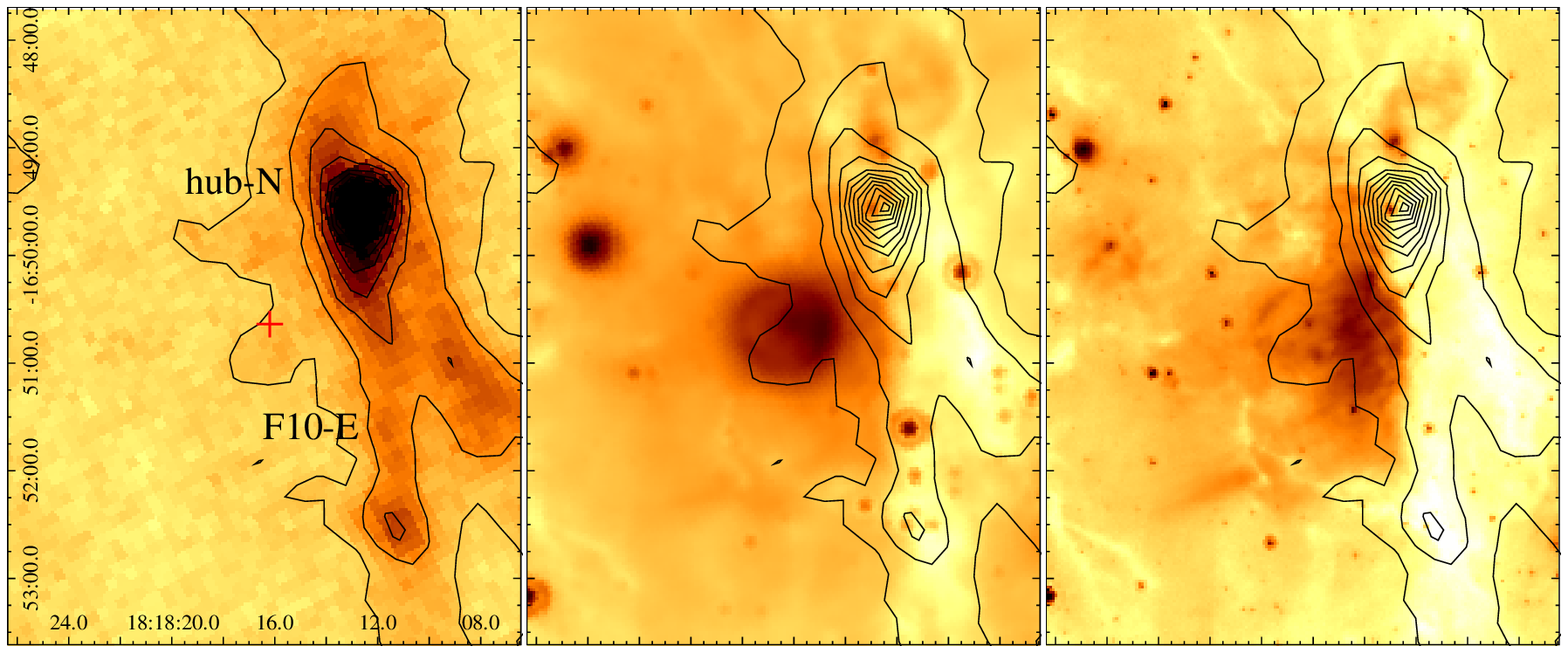}
\end{center}
\caption{Left: ATLASGAL 870\,$\mu$m image of the region around IRAS\,18153$-$1651, 
showing filamentary structures and a density enhancement (hub--N) at $\approx$1.5 
arcmin to the north-west from the position of stars\,1 and 2 (indicated by a red 
cross). Middle and right: {\it Spitzer} MIPS $24 \, \mu$m 
and IRAC 8\,$\mu$m images of the same region with the 870\,$\mu$m image overlayed in 
black contours. The coordinates are in units of RA (J2000) and Dec. (J2000) on the 
horizontal and vertical scales, respectively. At a distance of 2 kpc, 1 arcmin 
corresponds to $\approx$0.57 pc. See text for details.
    }
\label{fig:hub}
\end{figure*}

The nebula, which is the subject of this paper, was serendipitously discovered in the archival 
data of the {\it Spitzer Space Telescope} during our search for bow shocks generated by OB 
stars running away from the young massive star clusters NGC\,6611 and NGC\,6618 embedded in the 
giant \hii regions M16 and M17, respectively (for motivation and some results of this search, see 
Gvaramadze \& Bomans 2008; cf. Gvaramadze et al. 2014a). The {\it Spitzer} data we used were 
obtained with the Multiband Imaging Photometer for {\it Spitzer} (MIPS; Rieke et al. 2004) within 
the framework of the 24 and 70 Micron Survey of the Inner Galactic Disk with MIPS (Carey et al. 
2009) and with the {\it Spitzer} Infrared Array Camera (IRAC; Fazio et al. 2004) within the 
Galactic Legacy Infrared Mid-Plane Survey Extraordinaire (Benjamin et al. 2003). 

In the MIPS 24\,$\mu$m image the nebula appears as a limb-brightened, almost circular shell of 
radius $\approx$27 arcsec, with the western edge brighter than the other parts of the 
shell (see Fig.\,\ref{fig:neb}, upper left panel). This image also shows the presence of a point-like 
source, which is somewhat offset from the geometric centre of the shell in the north-west direction.
In the SIMBAD data base the nebula is named IRAS\,18153$-$1651, and we will use this name hereafter. 
At the distance of IRAS\,18153$-$1651 of 2 kpc (see below), 1 arcsec corresponds to $\approx$0.01 pc, 
so that the linear radius of the shell is $\approx$0.26 pc.

IRAS\,18153$-$1651 and the point source within it are also visible in all (3.6, 4.5, 5.8 and $8.0\,\mu$m) 
IRAC images (Fig.\,\ref{fig:neb}). In these images, the nebula lacks the circular shape and has a more
complicated appearance. At 8\,$\mu$m one still can see a more or less circular shell, whose west side  
overlaps with a bar-like structure stretched in the north-south direction. At shorter wavelengths, both 
the shell and the ``bar" become more diffuse. This morphology and the brightness asymmetry of the shell 
at 24\,$\mu$m indicate that the nebula is interacting with a dense medium in the west. This inference is 
supported by the data of the APEX Telescope Large Area Survey of the Galaxy (ATLASGAL; Schuller et al. 
2009). Fig.\,\ref{fig:hub} presents the ATLASGAL 870\,$\mu$m image of the region centred on 
IRAS\,18153$-$1651 (left-hand panel) and the {\it Spitzer} 24 and 8\,$\mu$m images of the same region 
(middle and right-hand panels, respectively) with the 870\,$\mu$m image overlayed in black contours. In 
the ATLASGAL image one can see two filaments of cold dense gas intersecting each other in a nodal point 
(dubbed ``hub--N" in Busquet et al. 2013; see below for more detail), while comparison of this image with 
the two other ones shows that IRAS\,18153$-$1651 is apparently interacting with the hub--N and one of 
the filaments (see also below).

\begin{figure}
\begin{center}
\includegraphics[width=8cm,angle=0]{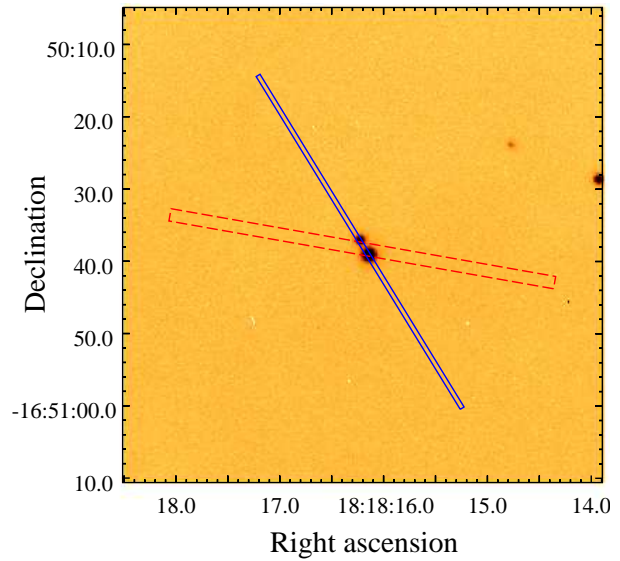}
\end{center}
\caption{GMOS $g\prime$-band acquisition image of the central stars of 
IRAS\,18153$-$1651, separated from each other by $\approx$2.3 arcsec or 
$\approx$0.02 pc in projection. The brightest of the stars is star\,1. 
The relative positions of the GMOS and TWIN slits are shown by a solid 
(blue) and a dashed (red) rectangle, respectively. The widths of the 
rectangles of 0.75 and 2.1 arcsec correspond to the widths of the slits.
    }
\label{fig:acq}
\end{figure}

The point source in IRAS\,18153$-$1651 can also be seen in all ($J,H,K_{\rm s}$) Two-Micron All Sky Survey 
(2MASS) images (Skrutskie et al., 2006) as well as in the images provided by the Digitized Sky Survey II 
(DSS-II) (McLean et al. 2000). The IRAC images clearly show that this source is composed of two nearby 
stars (see also Fig.\,\ref{fig:acq}). The coordinates of these stars, as given in the GLIMPSE Source Catalog 
(I + II + 3D) (Spitzer Science Center 2009), are: RA(J2000)=$18^{\rm h} 18^{\rm m} 16\fs21$, 
Dec.(J2000)=$-16\degr 50\arcmin 38\farcs8$ (hereafter star\,1) and RA(J2000)=$18^{\rm h} 18^{\rm m} 
16\fs28$, Dec.(J2000)=$-16\degr 50\arcmin 36\farcs7$ (hereafter star\,2). This catalogue also provides the 
IRAC band magnitudes for stars\,1 and 2 separately. The two stars are also resolved by the the UKIDSS Galactic 
Plane Survey (Lucas et al. 2008), which gives for them $H$- and $K$-band magnitudes. The total $B$- and 
$V$-band magnitudes of the two stars are, respectively, 15.84$\pm$0.05 and 14.14$\pm$0.05 (Henden et al. 
2016).	The details of the stars are summarized in Table\,\ref{tab:det}, to which we also added their spectral
types, effective temperatures and surface gravities, based on our spectroscopic observations and spectral 
analysis (presented in Sections\,\ref{sec:obs} and \ref{sec:sta}, respectively).

\begin{table}
  \caption{Details of two stars in the centre of IRAS\,18153$-$1651.
  The spectral types, SpT, effective temperatures, $T_{\rm eff}$ and surface gravities,
  $\log g$, are based on our spectroscopic observations and spectral analysis. 
  The coordinates and IRAC photometry are from the GLIMPSE Source Catalog 
  (I + II + 3D) (Spitzer Science Center 2009). The $H$ and $K$ photometry is 
  from Lucas et al. (2008). 
  }
  \label{tab:det}
  \renewcommand{\footnoterule}{}
  \begin{center}
  \begin{minipage}{\textwidth}
    \begin{tabular}{lcc}
      \hline
       & star\,1 & star\,2 \\
      \hline
      SpT & B1\,V & B3\,V \\
      $\alpha$ (J2000) & $18^{\rm h} 18^{\rm m} 16\fs21$ & $18^{\rm h} 18^{\rm m} 16\fs28$ \\
      $\delta$ (J2000) & $-16\degr 50\arcmin 38\farcs8$ & $-16\degr 50\arcmin 36\farcs7$ \\
      $H$ (mag) & 10.28$\pm$0.02 & 11.16$\pm$0.02 \\
      $K$ (mag) & 9.39$\pm$0.02 & 10.89$\pm$0.02 \\
      $[3.6]$ (mag) & 9.00$\pm$0.14 & --- \\
      $[4.5]$ (mag) & 8.98$\pm$0.15 & --- \\
      $[5.8]$ (mag) & 8.83$\pm$0.08 & 10.14$\pm$0.21 \\
      $T_{\rm eff}$ (kK) & 22$\pm$2 & 20$\pm$2 \\
      $\log g$ & 4.2$\pm$0.2 & 4.4$\pm$0.2 \\
      \hline
    \end{tabular}
    \end{minipage}
    \end{center}
    \end{table}

The DSS-II images of IRAS\,18153$-$1651 show diffuse emission to the west of stars\,1 and 2, extending 
to the edge of the 24\,$\mu$m shell. This emission is also clearly seen in the H$\alpha$+[N\,{\sc ii}] 
image (see Fig.\,\ref{fig:neb}) obtained in the framework of the SuperCOSMOS H-alpha Survey (SHS; Parker 
et al. 2005), which also reveals a clear arcuate structure located at about 12 arcsec (or 0.11 pc in 
projection) from the stars (note that these stars are not in the geometric centre of the arc; see 
Section\,\ref{sec:com} for discussion of this issue). The orientation of the arc suggests that it could 
be shaped by the winds of the central stars and that is what has motivated us to carry out the research 
presented in this paper.

A literature search showed that the region containing IRAS\,18153$-$1651 has been studied quite extensively 
during the last years (Busquet et a. 2013, 2016; Povich et al. 2016; Santos et al. 2016). These studies, 
however, only briefly touch IRAS\,18153$-$1651 and are mostly devoted to environments of this object. 
Below we review some relevant information about IRAS\,18153$-$1651.

IRAS\,18153$-$1651 is a part of the infrared dark cloud G14.225-0.506 (identified with {\it Spitzer} by 
Peretto \& Fuller 2009), which, in turn, is a central region of the south-west extension of the massive 
star-forming region M17 (M17\,SWex; Povich \& Whitney 2010). Using {\it Spitzer} and 2MASS data, Povich 
\& Whitney (2010) concluded that M17\,SWex is a precursor to an OB association and 
that this cloud as a whole will produce $>200$ B stars and perhaps not (m)any O stars. 
{\it Chandra} observations of G14.225-0.506 confirmed that M17\,SWex currently hosts no O-type stars 
(Povich et al. 2016). Observational evidence suggests that M17\,SWex is located in the Carina-Sagittarius 
arm at a distance of 2\,kpc (Povich et al. 2016). In what follows, we adopt this distance
for IRAS\,18153$-$1651 as well.

Radio observations of the dense NH$_3$ gas in G14.225-0.506 by Busquet et al. (2013) showed that this cloud 
consists of a net of eight molecular filaments, some of which intersect with each other in two regions of 
enhanced density (named hub--N and hub--S).
IRAS\,18153$-$1651 is located at about 1.5 arcmin south-east from one of these two
nodal points (hub--N) and is bounded from the west side by a 
filament (named F10--E) stretching for $\approx$2.5 arcmin from the hub--N to the south 
(see fig.\,2 in Busquet et al. 2013 and Fig.\,\ref{fig:hub}).
Busquet et al. (2013) attributed the origin of the filaments to the gravitational instability
of a thin layer threaded by a magnetic field. 
Polarimetric observations of background stars at optical and near-infrared wavelengths 
showed that the magnetic field lines in G14.225-0.506 are perpendicular to most of
the filaments and to the (elongated) cloud as a whole (Santos et al. 2016), which further supports the 
possibility that the regular
magnetic field plays an important role in the formation of parallel filaments observed 
in M17\,SWex and other star-forming regions. Interestingly, Santos et al. (2016) found 
that the magnetic field lines are not perpendicular to the filament F10--E and hub--N 
(which is also elongated in the north-south direction) and 
suggested that this discrepant behaviour might be caused by expansion of the \hii region 
IRAS\,18153$-$1651. Additional evidence suggesting the interaction between 
IRAS\,18153$-$1651 and the filament F10--E and hub--N was presented in Busquet et 
al. (2013, 2016).

The detection of numerous young stellar objects (Povich \& Whitney 2010) and H$_2$O masers 
(Jaffe, G\"usten \& Downes 1981; Wang et al. 2006) in M17\,SWex (including  G14.225-0.506) points to 
ongoing star formation in this region, while the presence of several bright IR sources, of which 
IRAS\,18153$-$1651 (with its bolometric luminosity of $\sim10^4 \, \lsun$ and number of Lyman 
continuum photons per second of $<1.5\times10^{46}$; Jaffe, Stier \& Fazio 1982)
is one of the brightest, implies that a number of massive stars have already formed 
there. {\it Chandra X-ray Observatory} imaging study of M17\,SWex revealed that IRAS\,18153$-$1651 contains one of the two
richest concentrations of X-ray sources detected in this star-forming region (Povich et al. 2016). Povich
et al. (2016) suggested that IRAS\,18153$-$1651 might represent a site for massive cluster formation,
and noted that the infrared and radio continuum luminosities of this \hii region indicate that it is powered by 
a B1-1.5\,V star. Similarly, Busquet et al. (2016) mentioned unpublished VLA 6cm observations of
IRAS\,18153$-$1651, which reveal ``a cometary \hii region ionized by a B1 star with the head of the 
cometary arch pointing toward hub--N". 
Our spectroscopic observations of stars\,1 and 2 confirm that IRAS\,18153$-$1651 is powered by 
a B1\,V star (see the next section) and suggest that this \hii region indeed contains a recently
formed star cluster (see Section\,\ref{sec:arc}).

\section{Spectroscopic observations}
\label{sec:obs}

To classify the central stars of IRAS\,18153$-$1651 and clarify 
the nature of the optical arc, we obtained 
long-slit spectra with the Gemini Multi-Object Spectrograph South
(GMOS-S) and the Cassegrain Twin Spectrograph (TWIN)
of the 3.5-m telescope in the Observatory of Calar Alto (Spain).

\subsection{Gemini-South}
\label{sec:GS}

To obtain spectra of stars\,1 and 2, we used the Poor Weather time 
at Gemini-South under the programme ID GS-2011B-Q-92. GMOS-S
provided coverage from 3800 to 6750\,\AA\, with a resolving power of 
$\approx$3000. The spectra were collected on 2012 May 7 under good seeing
conditions ($\approx$0.6 arcsec), but some thick clouds (extinction
around 1\,mag). The slit of width of 0.75 arcsec was aligned 
along stars\,1 and 2, i.e.
with a position angle (PA) of PA=$31\fdg5$, measured from north 
to east. The desired signal-to-noise ratio of $\sim$150 was
achieved with a total exposure time of 9$\times$150\,s. The bias
subtraction, flat-fielding, wavelength calibration and sky
subtraction were executed with the GMOS package in the Gemini
library of the {\sc iraf}\footnote{{\sc iraf}: the Image Reduction and
Analysis Facility is distributed by the National Optical Astronomy
Observatory (NOAO), which is operated by the Association of
Universities for Research in Astronomy, Inc. (AURA) under
cooperative agreement with the National Science Foundation (NSF).}
software. In order to fill the gaps in GMOS-S's CCD, the
observation was divided into three series of exposures obtained
with a different central wavelength, i.e. with a 5\,\AA \, shift 
between each exposure. The extracted spectra were obtained by averaging
the individual exposures, using a sigma clipping algorithm to
eliminate the effects of cosmic rays. The average wavelength
resolution is $\approx$0.46\,\AA\, pixel$^{-1}$ (full width at
half-maximum (FWHM) $\approx$4.09\,\AA), and the accuracy of the wavelength
calibration estimated by measuring the wavelength of 10 lamp
emission lines is 0.061\,\AA. A spectrum of the white dwarf H600 was
used for flux calibration and for removing the instrument response.
Unfortunately, due to the weather conditions, any absolute
measurement of the flux is not possible.

\subsection{Calar Alto}
\label{sec:CA}

An additional spectrum of IRAS\,18153$-$1651 was obtained
with TWIN on 2012 July 12 under the programme ID H12-3.5-013.
Three exposures of 600\,s were taken.
The set-up used for TWIN consisted of the gratings T08 in the first order
for the blue arm (spectral range 3500--5600 \AA) and T04 in the
first order for the red arm (spectral range 5300--7600 \AA) which
provide a reciprocal dispersion of 72~\AA\,${\rm mm}^{-1}$ for 
both arms. The resulting FWHM spectral resolution measured on
strong lines of the night sky and reference spectra was 3.1--3.7~\AA.
The Calar Alto observation was mostly intended to get a spectrum of the 
optical arc. Correspondingly, the slit of $240\times2.1$ arcsec$^2$ 
was oriented at PA=80$\degr$ to cross the brightest part of the 
arc. The seeing was variable, 
ranging from $\simeq$1.5 to 2.0 arcsec. Spectra of He--Ar 
comparison arcs were obtained to calibrate the wavelength scale 
and the spectrophotometric standard star BD+$33\degr$\,2642
(Oke 1990) was observed at the beginning of the night for
flux calibration.

The primary data reduction was done using {\sc iraf}: the data for 
each CCD detector were trimmed, bias subtracted and flat corrected. 
The subsequent long-slit data reduction was carried out in the 
way described in Kniazev et al. (2008). The blue and red parts of the 
spectra were reduced independently for all three exposures, then
aligned along the rows using the IRAF {\sc apall} task, and finally summed up.

\section{Stars\,1 and 2: spectral classification and modelling}
\label{sec:sta}

\begin{figure*}
\begin{center}
\includegraphics[width=14cm,angle=0]{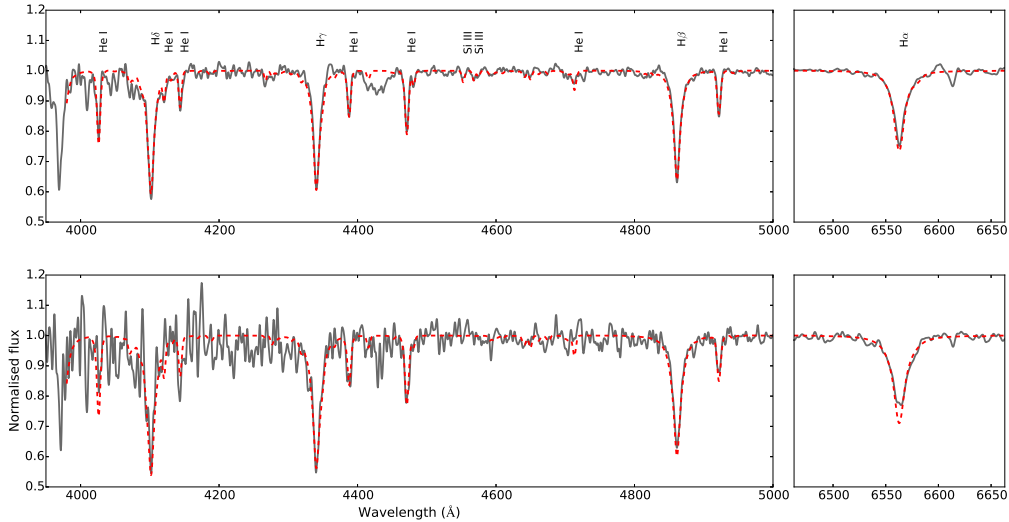}
\end{center}
\caption{Normalized spectra of stars\,1 and 2 observed with 
Gemini-South (plotted respectively in the upper and lower panels), 
compared with the best-fitting {\sc fastwind} model (red dashed line).
The lines fitted by the model are labelled. 
    }
\label{fig:mod}
\end{figure*}

To classify stars\,1 and 2, we used their GMOS spectra. The spectra 
are dominated by H and He\,{\sc i} absorption 
lines (see Fig.\,\ref{fig:mod}). No He\,{\sc ii} lines are visible in the 
spectra, which implies that both stars are of B type. Using the 
EW(H$\gamma$)--absolute magnitude calibration by Balona \& Crampton (1974) 
and the measured equivalent widths of EW(H$\gamma$)=4.5$\pm$0.1\,\AA \,
and 6.3$\pm$0.4\,\AA \, for stars\,1 and 2, respectively, we estimated 
their spectral types as B1\,V and B3\,V. The apparent luminosity
resulting from the B1\,V classification of star\,1 is consistent with the location 
of IRAS\,18153$-$1651 at the distance of 2 kpc (cf. Povich et al. 2016).
We note that asymmetric profiles of the He\,{\sc i} lines in the spectrum 
of star\,2 suggest that this star might be a binary system. Nonetheless, 
the low spectral resolution and signal-to-noise ratio did not allow 
us to confirm this.

We modelled the spectra using the approach described in Castro et al. (2012; 
see also Lefever et al. 2010). The technique is rooted in a previously calculated 
{\sc fastwind}\footnote{The stellar atmosphere code {\sc fastwind} provides 
reliable synthetic spectra of O- and B-type stars taking  into account non-local 
thermodynamic equilibrium effects in spherical symmetry with an explicit treatment 
of the stellar wind.} (Santolaya-Rey, Puls \& Herrero 1997; Puls et al. 2005) 
stellar atmosphere grid (e.g. Simon-Diaz et al. 2011; Castro et al. 2012) and a 
$\chi^{2}$-based algorithm, searching for the best set of parameters that reproduce 
the main strong optical lines observed between $\approx$4000$-$5000\,\AA \, (labelled 
in Fig.\,\ref{fig:mod}). With the best-fitting models for stars\,1 and 2 (see 
Fig.\,\ref{fig:mod}), we derived effective temperatures of 
$T_{\rm eff}$=22\,000$\pm$2\,000\,K and 20\,000$\pm$2\,000\,K, and surface gravities 
of $\log g$=4.2$\pm$0.2 and 4.4$\pm$0.2, respectively. 

Main-sequence stars of these effective temperatures are sources of radiatively driven 
winds, which are strong enough to appreciably modify the ambient medium. In 
Table\,\ref{tab:krt} we give mass-loss rates, $\dot{M}$, and terminal wind velocities,
$v_\infty$, predicted by Krti\v{c}ka (2014) for main-sequence B stars with $T_{\rm eff}$ 
in the range of temperatures derived for stars\,1 and 2, i.e. for 
$T_{\rm eff}\in[18\,000,24\,000]$\,K. In this table we also give the rates of Lyman and 
dissociating Lyman-Werner photons (respectively, $Q_0$ and $Q_{\rm FUV}$) for the same 
range of effective temperatures (taken from Diaz-Miller, Franco \& Shore 1998). The 
spectral classification of stars\,1 and 2 implies that $T_{\rm eff}$ of star\,1 should 
be at the upper end of the temperature range derived from spectral modelling, while that 
of star\,2 -- at the lower end (cf. Kenyon \& Hartmann 1995). In the following, we 
adopt for these stars the effective 
temperatures of 24 and 18 kK, respectively. From Table\,\ref{tab:krt} it then follows 
that the mechanical wind luminosity,  $L_{\rm w}=\dot{M}v_{\infty}^2 /2$, and radiation 
fluxes of star\,1 are more than two orders of magnitude higher then those of star\,2, so 
that we will consider the former star as the main energy source in IRAS\,18153$-$1651.

\begin{table}
\caption{Mass-loss rates and terminal wind velocities of main-sequence 
B stars of solar metallicity and different effective temperatures as 
predicted by Krti\v{c}ka (2014), and rates of Lyman ($Q_0$) and dissociating 
Lyman-Werner ($Q_{\rm FUV}$) photons from Diaz-Miller et al. (1998).} 
\label{tab:krt}
\begin{center}
\begin{minipage}{\textwidth}
\begin{tabular}{ccccc}
\hline
$T_{\rm eff}$ & $\dot{M}$ & $v_\infty$ & $Q_0$ & $Q_{\rm FUV}$ \\
(kK) & ($\myr$) & ($\kms$) & (${\rm s}^{-1}$) & (${\rm s}^{-1}$) \\
\hline
18 & $9.1\times10^{-12}$ & 820    & $1.86\times10^{43}$ & $2.04\times10^{46}$ \\
20 & $3.4\times10^{-11}$ & 1\,290 & $1.35\times10^{44}$ & $6.03\times10^{46}$ \\
22 & $7.9\times10^{-11}$ & 1\,690 & $6.31\times10^{44}$ & $1.38\times10^{47}$ \\
24 & $3.9\times10^{-10}$ & 1\,700 & $2.40\times10^{45}$ & $2.45\times10^{47}$ \\
\hline
\end{tabular}
\end{minipage}
\end{center}
\end{table}

\section{Optical arc}
\label{sec:arc}

The GMOS slit was oriented along stars\,1 and 2 and therefore it does not cross 
the optical arc. Correspondingly, we did not detect any signatures of nebular 
emission in the 2D spectrum. On the contrary, the slit of the TWIN spectrograph 
was oriented in such a way that it intersects the region of maximum brightness 
of the arc. Below we discuss the spectrum of the arc obtained with this spectrograph.

\begin{figure*}
\begin{center}
\includegraphics[width=15cm,angle=0,clip=]{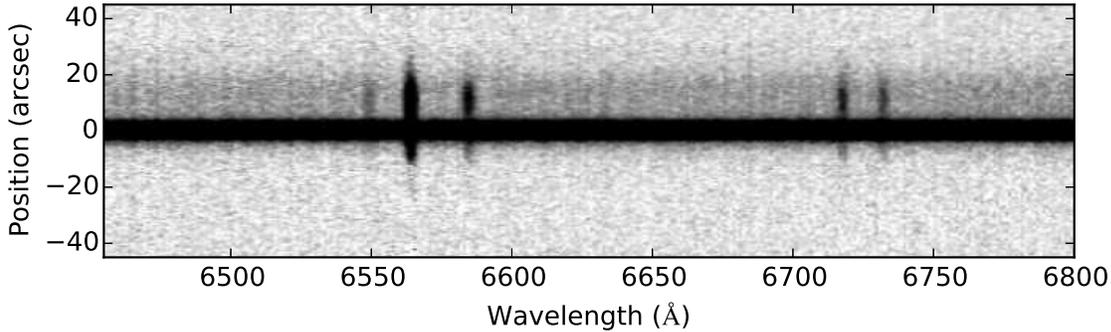}
\end{center}
\caption{A part of the TWIN 2D spectrum of 
the optical emission around stars\,1
and 2, showing nebular emission lines (from left to right) of [N\,{\sc ii}] 
$\lambda$6548, H$\alpha$, [N\,{\sc ii}] $\lambda$6584 and [S\,{\sc ii}] 
$\lambda\lambda$6717, 6731. The upper part of the spectrum corresponds to 
the emission west of the stars. Note the presence of a strong continuum 
emission in this direction.} 
\label{fig:2D}
\end{figure*}

\begin{figure}
\begin{center}
\includegraphics[width=8cm,angle=0,clip=]{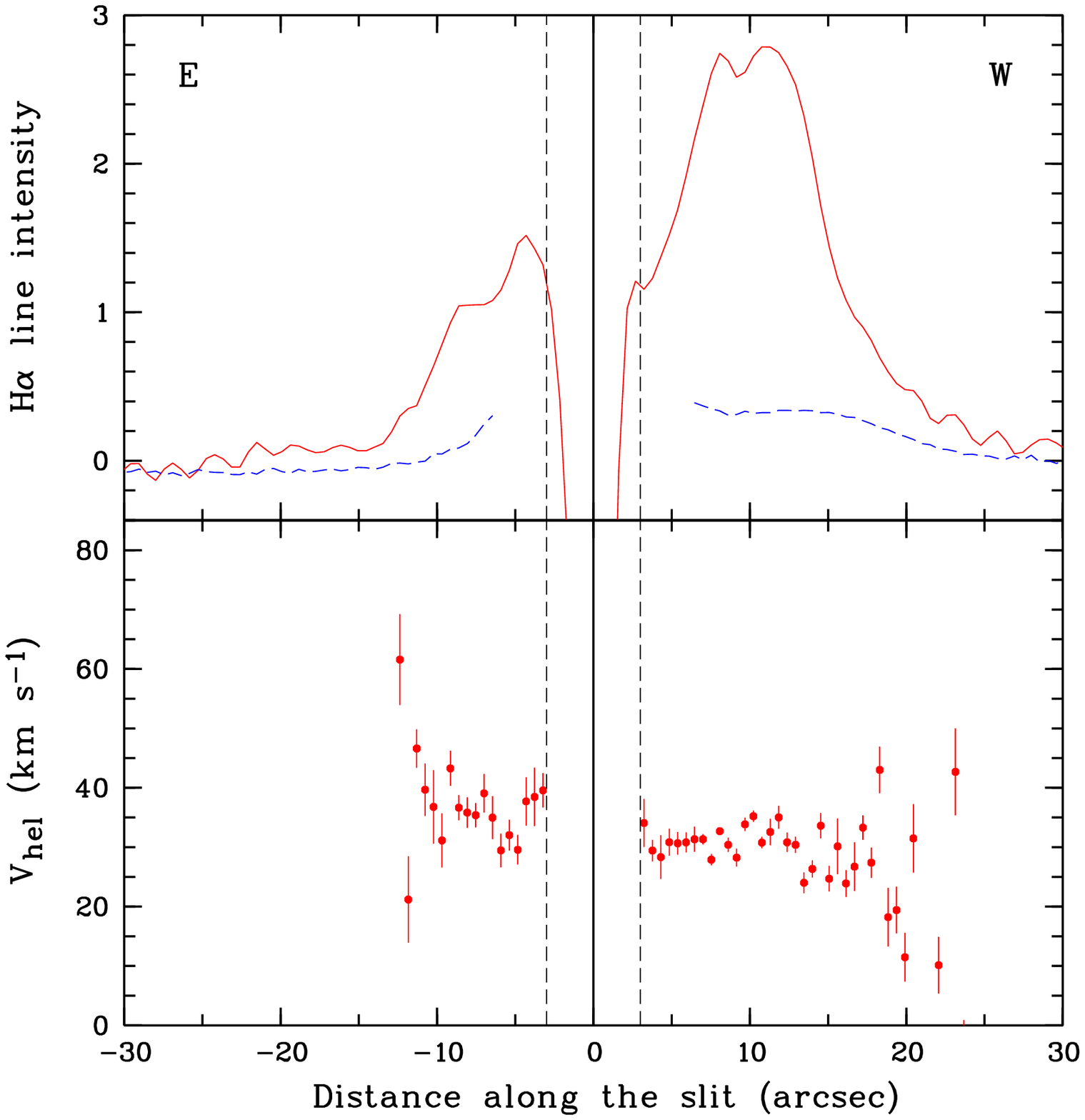}
\includegraphics[width=8cm,angle=0,clip=]{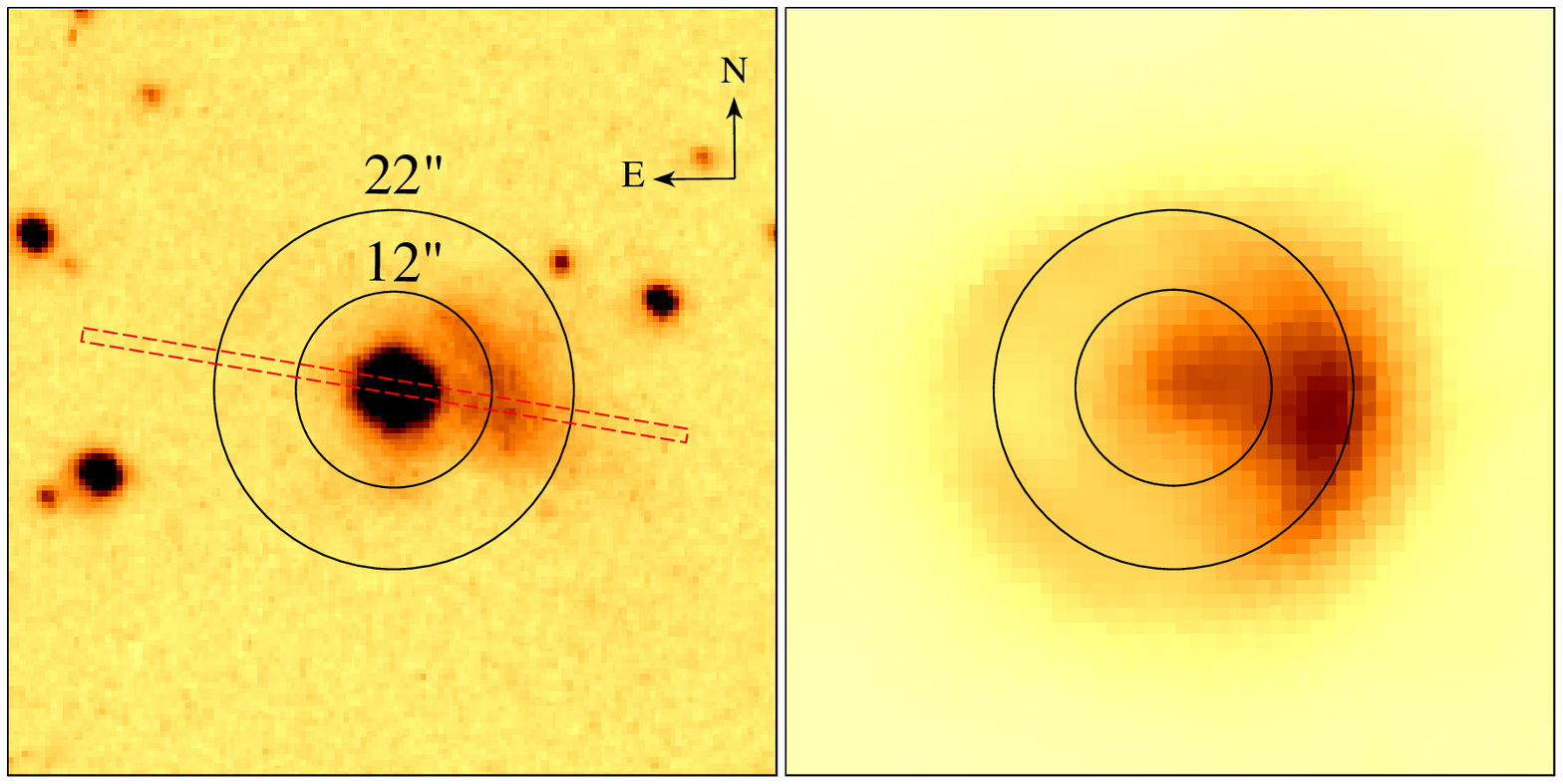}
\end{center}
\caption{Upper panel: H$\alpha$ line and continuum emission intensities along the TWIN slit 
(PA=80$\degr$), shown respectively with a solid (red) and dashed (blue) lines. Middle panel:
H$\alpha$ heliocentric radial velocity distribution along the slit. The E--W direction of 
the slit is shown. The solid vertical line corresponds to the position of star\,1, while 
the dashed vertical lines (at $\pm$3 arcsec from the solid one) mark the area where the radial 
velocity was not measured because of the effect of stars\,1 and 2. Bottom panel: 
SHS H$\alpha$+[N\,{\sc ii}] (left) and MIPS $24 \, \mu$m (right) images of 
IRAS\,18153$-$1651 with the position of the TWIN slit shown by a red dashed rectangle. 
Concentric circles of radii 12 and 22 arcsec are overplotted on the images to make their 
comparison with the upper panels more convenient.} 
\label{fig:Ha}
\end{figure}

In the TWIN 2D spectrum the arc becomes visible via its emission lines of H$\alpha$, H$\beta$,
[N\,{\sc ii}] $\lambda\lambda$6548, 6584 and [S\,{\sc ii}] $\lambda\lambda$6717, 6731 
superimposed on strong continuum emission, which extends to the west of stars\,1 
and 2 for about 20 arcsec (see also Fig.\,\ref{fig:Ha}). A part of this spectrum 
is presented in Fig.\,\ref{fig:2D}. Such composed spectra are typical of ionized 
reflection nebulae like the Orion Nebula (see e.g. plate XXIX in Greenstein \& 
Henyey 1939). Correspondingly, we attribute the continuum emission to the starlight 
scattered by dust in the shell around the \hii region.

\begin{figure}
\begin{center}
\includegraphics[angle=270,width=1.0\columnwidth,clip=]{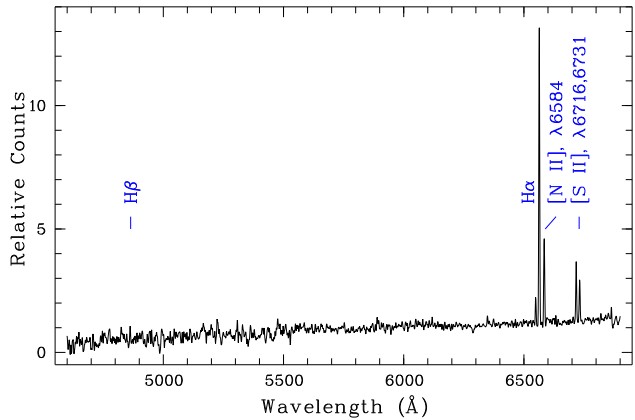}
\end{center}
\caption{1D spectrum of the optical arc obtained with the TWIN spectrograph.} 
\label{fig:neb-spec}
\end{figure}

We extracted a 1D spectrum over the optical arc by summing up, without any weighting, 
all rows from the area of an annulus with an outer radius of 20 arcsec centred on 
stars\,1 and 2 and the central $\pm$3 arcsec excluded. The resulting spectrum is 
presented in Fig.\,\ref{fig:neb-spec}. The
emission lines detected in the spectrum were measured using the
programs described in Kniazev et al. (2004). Table\,\ref{tab:int}
lists the observed intensities of these lines normalized to
H$\beta$, $F$($\lambda$)/$F$(H$\beta$), the reddening-corrected
line intensity ratios, $I$($\lambda$)/$I$(H$\beta$), and the logarithmic
extinction coefficient, $C$(H$\beta$), which corresponds to a colour 
excess of $E(B-V)$=2.08$\pm$0.23 mag. In Table\,\ref{tab:int} we also give
the electron number density derived from the intensity ratio of
the [S\,{\sc ii}] $\lambda\lambda$6716, 6731 lines, $n_{\rm
e}$([S\,{\sc ii}]). Both $C$(H$\beta$) and $n_{\rm e}$([S\,{\sc
ii}]) were calculated under the assumption that $T_{\rm e}=6\,500$ K 
(see Section\,\ref{sec:sim}). All calculations were done in the 
way described in detail in Kniazev et al. (2008).

\begin{table}
\centering{\caption{Line intensities in the Calar Alto spectrum of
the optical arc.} \label{tab:int}
\begin{tabular}{lll} \hline
\rule{0pt}{10pt}
\rule{0pt}{10pt}
$\lambda_{0}$(\AA) Ion & $F$($\lambda$)/$F$(H$\beta$)&$I$($\lambda$)/$I$(H$\beta$) \\ 
\hline
4861\ H$\beta$\            & 1.00$\pm$0.37 & 1.00$\pm$0.39 \\
6548\ [N\ {\sc ii}]\       & 2.28$\pm$0.64 & 0.22$\pm$0.07 \\
6563\ H$\alpha$\           &31.21$\pm$8.12 & 2.94$\pm$0.87 \\
6584\ [N\ {\sc ii}]\       & 9.07$\pm$2.38 & 0.83$\pm$0.25 \\
6716\ [S\ {\sc ii}]\       & 6.07$\pm$1.59 & 0.48$\pm$0.07 \\
6731\ [S\ {\sc ii}]\       & 4.23$\pm$1.11 & 0.33$\pm$0.05 \\
  & & \\
$C$(H$\beta$)             & \MC {2}{l}{3.06$\pm$0.34}  \\
$E(B$$-$$V$)         & \MC {2}{l}{2.08$\pm$0.23 mag}  \\
$n_{\rm e}$([S\,{\sc ii }]) & \MC {2}{l}{10$^{+550}_{-10}$ cm$^{-3}$}  \\
\hline
\end{tabular}
 }
\end{table}

The upper and middle panels in Fig.\,\ref{fig:Ha} plot, respectively, the H$\alpha$ line 
and continuum intensities and the H$\alpha$ heliocentric radial velocity distribution 
along the slit. A comparison of these panels with the SHS and MIPS images of IRAS\,18153$-$1651
(presented for convenience in the bottom panel of Fig.\,\ref{fig:Ha}) shows that in the west 
direction from stars\,1 and 2 the H$\alpha$ and 
continuum emission extend to the edge of the mid-infrared nebula, and that the 
H$\alpha$ intensity peaks at the position of the arc. In the opposite direction, 
the H$\alpha$ line intensity is prominent up to a distance comparable to the radius 
of the arc, while the continuum emission fades at a shorter distance. From the lower 
panel of Fig.\,\ref{fig:Ha}, we derived the mean heliocentric radial velocity of the 
H$\alpha$ emission of $31\pm5 \, \kms$. Using this velocity and assuming the distance 
to the Galactic Centre of $R_0$=8.0 kpc and the circular rotation speed of the Galaxy 
of $\Theta _0 =240 \, \kms$ (Reid et al. 2009), and the solar peculiar motion
$(U_{\odot},V_{\odot},W_{\odot})=(11.1,12.2,7.3) \, \kms$ (Sch\"onrich, Binney \& 
Dehnen 2010), one finds the local standard of rest velocity of $27\pm5 \, \kms$, 
which agrees well with that of the cloud G14.225-0.506 (Jaffe et al. 1982; Busquet
et al. 2013) and the star-forming region M17\,SWex as a whole (e.g. Povich et al. 
2016).

The electron number density could also be derived from the surface
brightness of the arc in the H$\alpha$ line, $S_{{\rm H}\alpha}$,
measured on the SHS image (cf. equation 4 in Frew et al. 2014):
\begin{eqnarray}
n_{\rm e}\approx 1.7 \, {\rm cm}^{-3} \left({l\over 1 \, {\rm
pc}}\right)^{-0.5}\left({T_{\rm e} \over 10^4 \, {\rm K}}\right)^{0.45} \nonumber \\
\times \left({S_{{\rm H}\alpha}\over 1 \, {\rm
R}}\right)^{0.5}e^{1.1E(B-V)} \, ,
\label{eqn:den}
\end{eqnarray}
where $l$ is the line-of-sight thickness of the arc and
1\,R$\equiv$1\,Rayleigh=5.66$\times10^{-18}$ erg cm$^{-2}$
s$^{-1}$ arcsec$^{-2}$ at H$\alpha$ (here we assumed that $n_{\rm
e}$ and $T_{\rm e}$ are constant within the arc).

Using equations\,(1) and (2) in Frew et al. (2014) and the flux
calibration factor of 20.4 counts pixel$^{-1}$ R$^{-1}$ from their
table\,1, and adopting the observed [N\,{\sc ii}] to H$\alpha$
line intensity ratio of $0.36^{+0.30} _{-0.17}$ from Table\,\ref{tab:int} 
(here [N\,{\sc ii}] corresponds to the sum of the $\lambda$6548 and
$\lambda$6584 lines), we obtained the peak surface brightness of
the arc (corrected for the contribution from the contaminant 
[N\,{\sc ii}] lines) of $S_{{\rm H}\alpha}\approx$30$^{+3} _{-4}$\,R. 
Then, using equation\,(\ref{eqn:den}) with $E(B-V)$=2.08$\pm$0.23 mag,
$l$$\approx$0.23 pc (we assumed that the arc is a part of
a spherical shell of inner radius of 12 arcsec and thickness 
of 5 arcsec) and $T_{\rm e}$=6\,500\,K (see Section\,\ref{sec:sim}), 
one finds $n_{\rm e}$$\approx$160$^{+45} _{-35}$ \,
${\rm cm}^{-3}$, which agrees within the error margins with 
$n_{\rm e}$([S\,{\sc ii}]) given in Table\,\ref{tab:int}.

To find the actual H$\alpha$ surface brightness of the optical arc, 
$S_{{\rm H}\alpha} ^{\rm act}$,
one needs to estimate the attenuation of the H$\alpha$ emission line 
in magnitudes, $A$(H$\alpha$), in the direction towards IRAS\,18153$-$1651, 
which is related to the visual extinction, $A_V=R_VE(B-V)$, through the 
following relationship: $A$(H$\alpha$)=$0.828A_V$ (e.g. James et al. 2005).
Assuming a ratio of total to selective extinction of 
$R_V=3.1$ and using $E(B-V)$ from Table\,\ref{tab:int}, one finds 
$A$(H$\alpha$)=5.34$\pm$0.59 mag and $S_{{\rm H}\alpha} 
^{\rm act}\approx4\,100^{+3\,700} _{-2\,100}$\,R.

Emission-line objects can be classified by using various diagnostic 
diagrams (see e.g. Kniazev, Pustilnik \& Zucker 2008;
Frew \& Parker 2010, and references therein), of which the most 
frequently used one is $\log$(H$\alpha$/([S\,{\sc ii}]
$\lambda\lambda$6716, 6731)) versus $\log$(H$\alpha$/([N\,{\sc ii}]
$\lambda\lambda$6548, 6584)). For the optical arc one has
$0.56^{+0.18} _{-0.20}$ versus 0.45$\pm$0.27 (see Table\,\ref{tab:int}). 
These values place the arc in the area occupied by \hii regions, although 
the large error bars 
allow the possibility that the arc is located within the domain 
occupied by supernova remnants, which means that the emission of the 
arc might be due to shock excitation. 

Proceeding from this, one can suppose that the arc is either the edge of 
an asymmetric stellar wind bubble, distorted by a density gradient and/or 
stellar motion (e.g. Mackey et al. 2015, 2016), or a bow shock if the 
relative velocity between the wind-blowing star and the ambient medium 
is higher than the sound speed of the latter.
The detection of two early type B stars separated from each 
other by only $\approx$2.3 arcsec (or 0.02 pc in projection) and the 
presence of a concentration of X-ray sources around them (see Section\,\ref{sec:bub}) 
suggest that we deal with a recently 
formed star cluster with stars\,1 and 2 being its most massive members
(cf. Gvaramadze et al. 2014b). This, in turn, implies that the mass of the cluster 
is about $100 \, \msun$ (e.g. Kroupa et al. 2013).
We hypothesize therefore that the arc is the edge of the wind bubble blown by the 
wind of star\,1 and suggest that the one-sided appearance of the bubble is caused by 
the interaction between the bubble and a photoevaporated flow from the molecular cloud 
to the west of the star (see Section\,\ref{sec:bub}). 

Further, we interpret the more extended (mid-infrared) nebula around the arc as 
an \hii region, so that the radius of the nebula is equal to the Str\"omgren radius 
$R_{\rm S}$. Proceeding from this, one can estimate the electron number density of the 
ambient medium:
\begin{equation}
n_{\rm e}=\sqrt{{3Q_0\over 4\pi\alpha _{\rm B}R_{\rm S}^3}} \, ,
\label{eqn:dens}
\end{equation}
where $\alpha _{\rm B} =3.4\times10^{-10}T^{-0.78}$ is the Case\,B recombination 
coefficient (Hummer 1994). This assumes that the electron and H$^+$ number densities
are the same, which is true for B stars because they cannot ionize helium. 
If $T$=6\,500\,K (see Section\,\ref{sec:sim}) and assuming $Q_0=2.4\times10^{45} 
\, {\rm s}^{-1}$ (see Table\,\ref{tab:krt}), one finds from equation\,(\ref{eqn:dens})
that $n_{\rm e}\sim100 \, {\rm cm}^{-3}$.

In such a dense medium the bubble created by the wind of star\,1 will soon become
radiative (i.e. the radiative losses of the shocked wind material at the interface
with the ISM become comparable to $L_{\rm w}$; cf. Mackey et al. 2015). This happens at 
the moment (McCray 1983):
\begin{equation}
t_{\rm rad}=1.5\times10^4 L_{33} ^{0.3} n_{100} ^{-0.7} \, {\rm yr} \, ,
\label{eqn:time}
\end{equation}
when the radius of the bubble is 
\begin{equation}
R_{\rm rad}=0.2L_{33} ^{0.4} n_{100} ^{-0.6} \, {\rm pc} \, ,
\label{eqn:rad}
\end{equation}
where $L_{33}=L_{\rm w}/(10^{33} \, {\rm erg} \, {\rm s}^{-1})$ and $n_{100}=n/(100 \, 
{\rm cm}^{-3})$. Subsequent evolution of the bubble follows the momentum-driven solution 
by Steigman et al. (1975) and its radius is given by:
\begin{equation}
R(t)=R_{\rm rad}(t/t_{\rm rad})^{1/2} \, .
\label{eqn:ste}
\end{equation}
With $\dot{M}=3.9\times10^{-10} \, \myr$ and $v_{\infty}=1700 \, \kms$ (see
Table\,\ref{tab:krt}), assuming $n=200 \, {\rm cm}^{-3}$ (see Section\,\ref{sec:sim}),  
and using equations\,(\ref{eqn:time}) and (\ref{eqn:rad}), one finds that $L_{\rm w}\approx3.6\times10^{32} 
\, {\rm erg} \, {\rm s}^{-1}$, $t_{\rm rad}\approx7\,000$\,yr and $R_{\rm rad}\approx0.09$\,pc. 
Then, it can be seen from equation\,(\ref{eqn:ste}) that the radius of the bubble would 
be equal to the observed radius of the optical arc of $\approx$0.11\,pc if the bubble 
was formed only $\approx11\,000$\,yr ago\footnote{We note that these estimates are very
approximate and should be considered with caution.}. This inference is supported by numerical modelling 
presented in the next section.

\section{Numerical Modelling}
\label{sec:num}
\subsection{Simulations}
\label{sec:sim}

To support our scenario for the origin of the optical arc, we ran 2D axisymmetric 
simulations using the \textsc{pion} code (Mackey \& Lim 2010; Mackey 2012). 
The calculations solve the Euler equations of hydrodynamics 
and are accurate to second order in space and time. In addition, the 
non-equilibrium ionization fraction of hydrogen and heating and cooling 
processes are also calculated, mediated by the ionizing (EUV) and non-ionizing 
UV (FUV) radiation field that is calculated by a raytracing scheme.
A uniform grid was used with cylindrical coordinates $z\in[-0.512,0.512]$\,pc 
and $R\in[0,0.512]$\,pc resolved by 1536 and 768 grid zones, respectively, for 
a grid resolution of $5\times10^{-4}$\,pc. Rotational symmetry around $R=0$ 
is assumed, and the simulations used the same general setup as Mackey et al. 
(2015).

We consider a single stellar source (a B1 star), with $T_{\rm eff}=24$\,kK, 
using the wind and radiation properties in Table\,\ref{tab:krt}:
$\dot{M}=4\times10^{-10} \myr$, $v_\infty=1700 \, \kms$, 
$Q_0=2.40\times10^{45}$\,s$^{-1}$ and $Q_{\rm FUV}=2.45\times10^{47}$\,s$^{-1}$.
The FUV radiation heats the neutral gas around the star through photoelectric 
heating on grains, following the implementation of Henney et al. (2009). 
A uniform ISM was set up around the star with a H number density of $n_{\rm H}$=100,
200 and 300\,cm$^{-3}$. In the following, however, we present only the 
results of simulations with $n_{\rm H}$=200\,cm$^{-3}$ (or a mass density of
$\rho=4.68\times10^{-22}$g\,cm$^{-3}$) because they better match the observations. 
At $z>0.1$\,pc the ISM was set to be 5 times denser, to mimic the nearby molecular cloud 
(although the real structure of the cloud must be more complicated, with substructure and 
density gradients; see Section\,\ref{sec:bub} and Fig.\,\ref{fig:hub}). The gas is initially 
at rest and in pressure equilibrium, with $p/k_{\rm B}=1.1\times10^4\,$K\,cm$^{-3}$,
where $p$ is the gas pressure and $k_{\rm B}$ is the Boltzmann constant.

\begin{figure}
\includegraphics[width=0.45\textwidth]{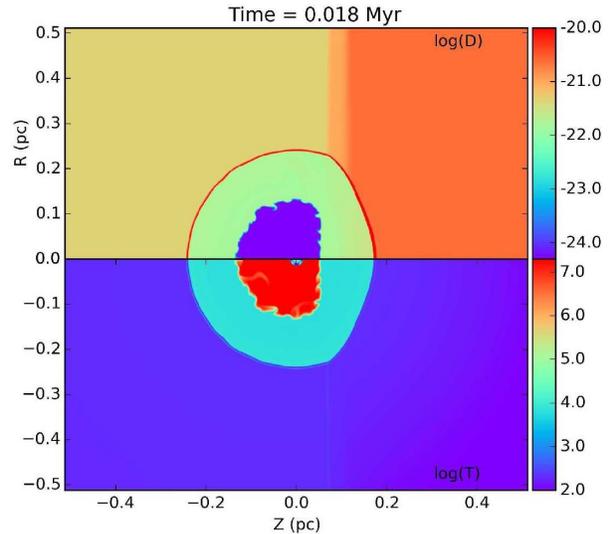}
\caption{Gas density (upper half-plane) and temperature (lower half-plane) for the 
2D simulation of the wind bubble and \hii region described in the text, after 0.018\,Myr 
of evolution. The units of density are $\log (\rho/{\rm g}\,{\rm cm}^{-3})$ and of 
temperature are $\log(T/{\rm K})$. This plot shows the simulated 2D plane, i.e., 
\emph{not} projected onto the plane of the sky.}
\label{fig:simsDT}
\end{figure}

The evolution of this simulation is now described. The \hii region expands rapidly 
in a spherical manner until it reaches the density jump at $z=0.1$\,pc.
At this interface a shock is transmitted into the dense medium and a second shock 
is reflected back into the \hii region. This reflected flow subsequently develops 
into a photoevaporation flow from the dense medium. The equilibrium temperature 
of the \hii region is $T\approx6500$\,K because of the very soft spectrum of the 
B1 star. The stellar wind drives a hot, expanding bubble within the \hii region that 
is initially spherical. Subsequently this bubble is impacted by the trans-sonic flow from the 
dense medium through the \hii region and is deformed. This creates a compressed and 
overpressurised layer between the dense molecular cloud and wind bubble. The ISM 
displaced by the wind bubble also creates an 
overdense layer in all directions at the interface between wind bubble and \hii region.
These two overdense regions are the brightest emitters in H$\alpha$. A snapshot from 
this simulation is shown in Fig.\,\ref{fig:simsDT}, where gas density and temperature 
are plotted on logarithmic scales.

\subsection{Synthetic maps and comparison with observations}
\label{sec:com}

\begin{figure*}
\centering
\includegraphics[width=0.33\textwidth]{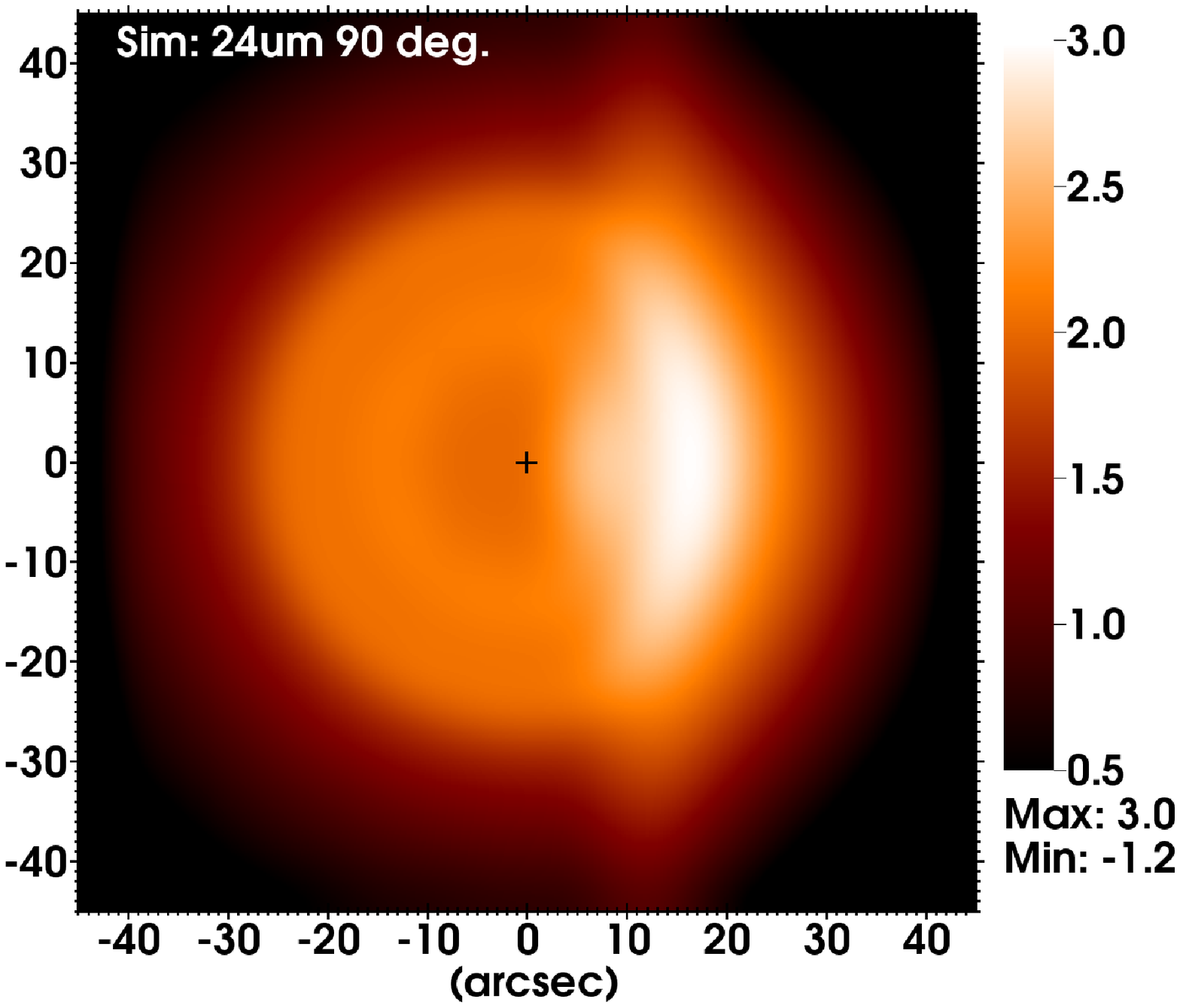}
\includegraphics[width=0.33\textwidth]{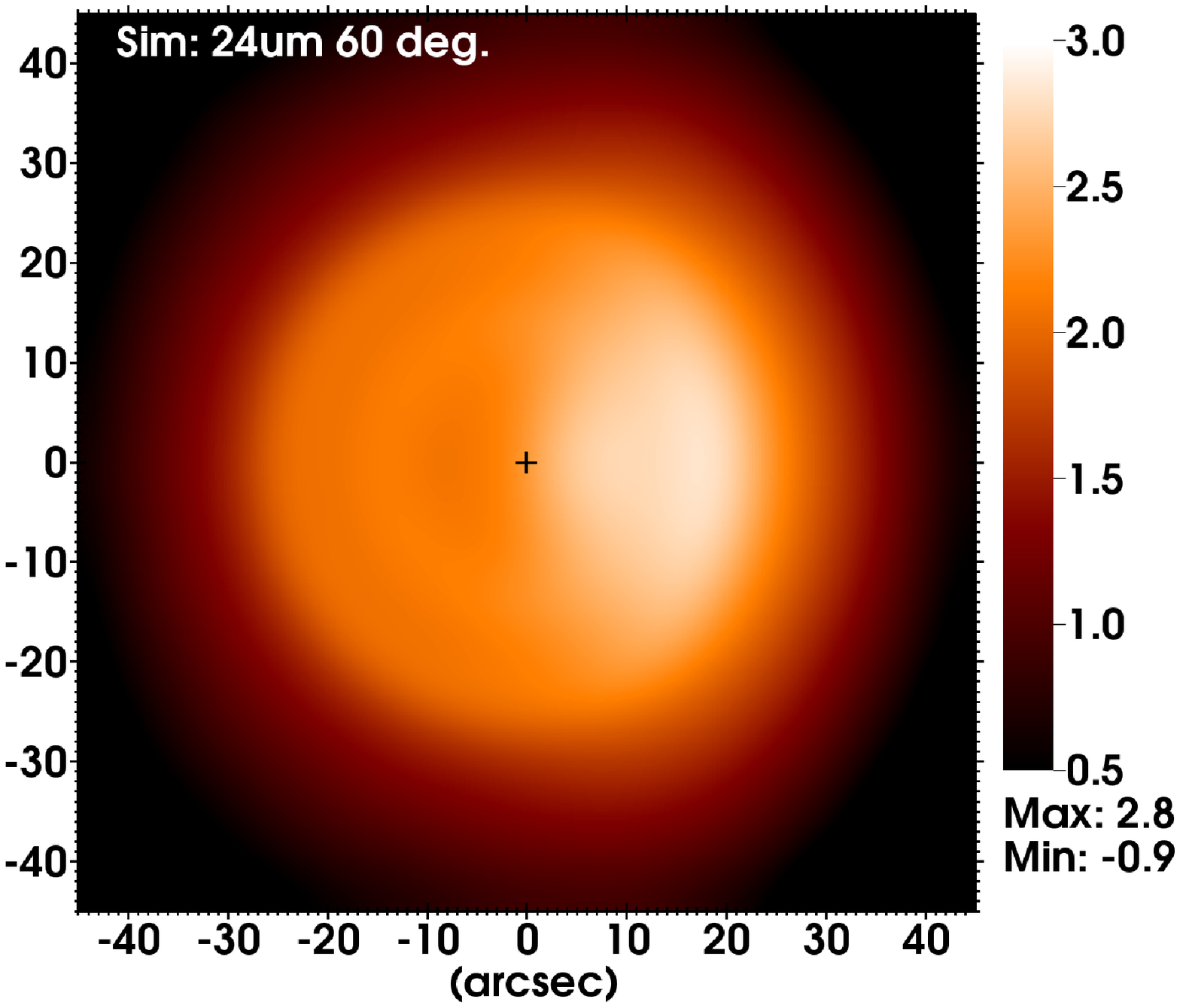}
\includegraphics[width=0.33\textwidth]{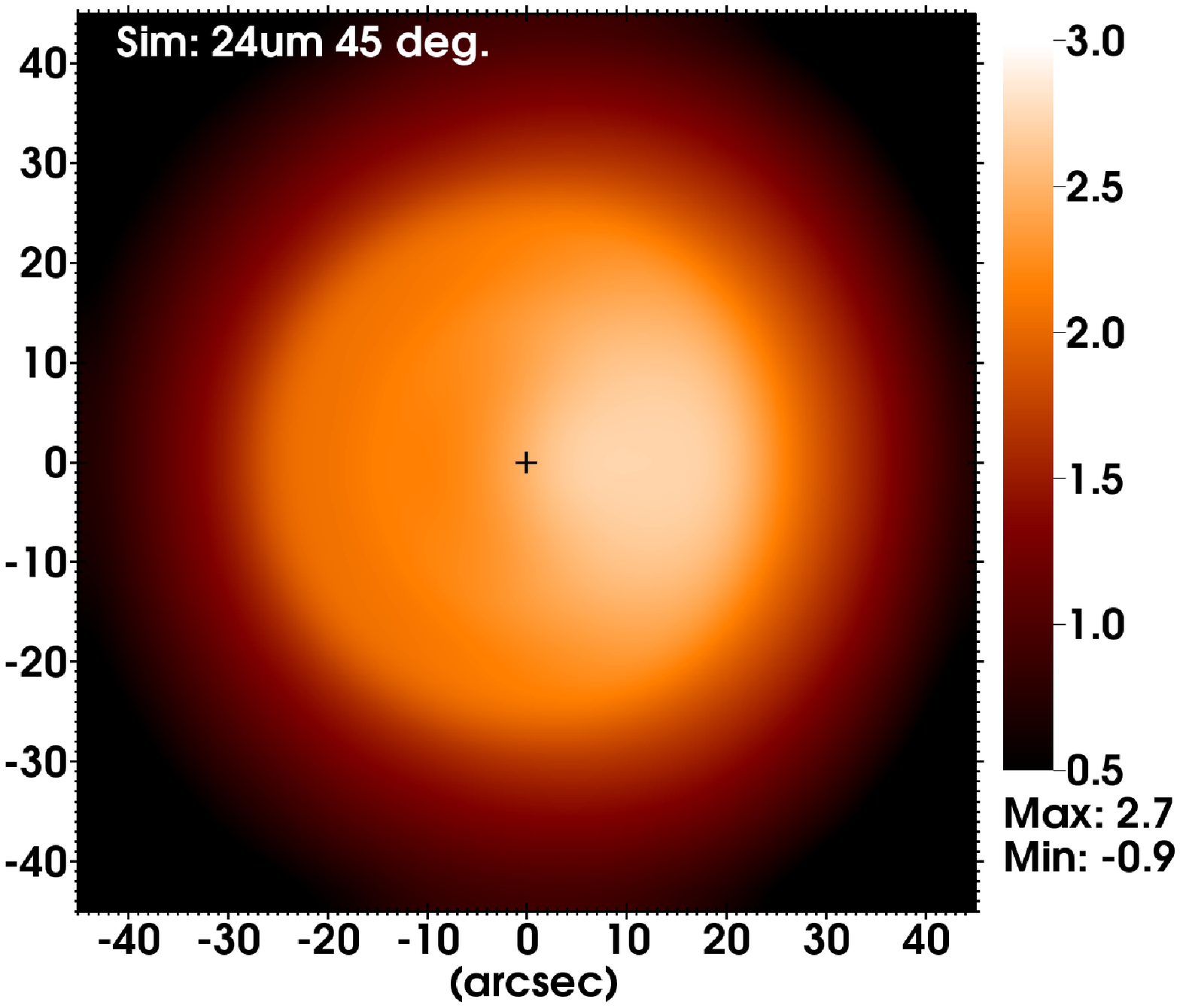}
\includegraphics[width=0.33\textwidth]{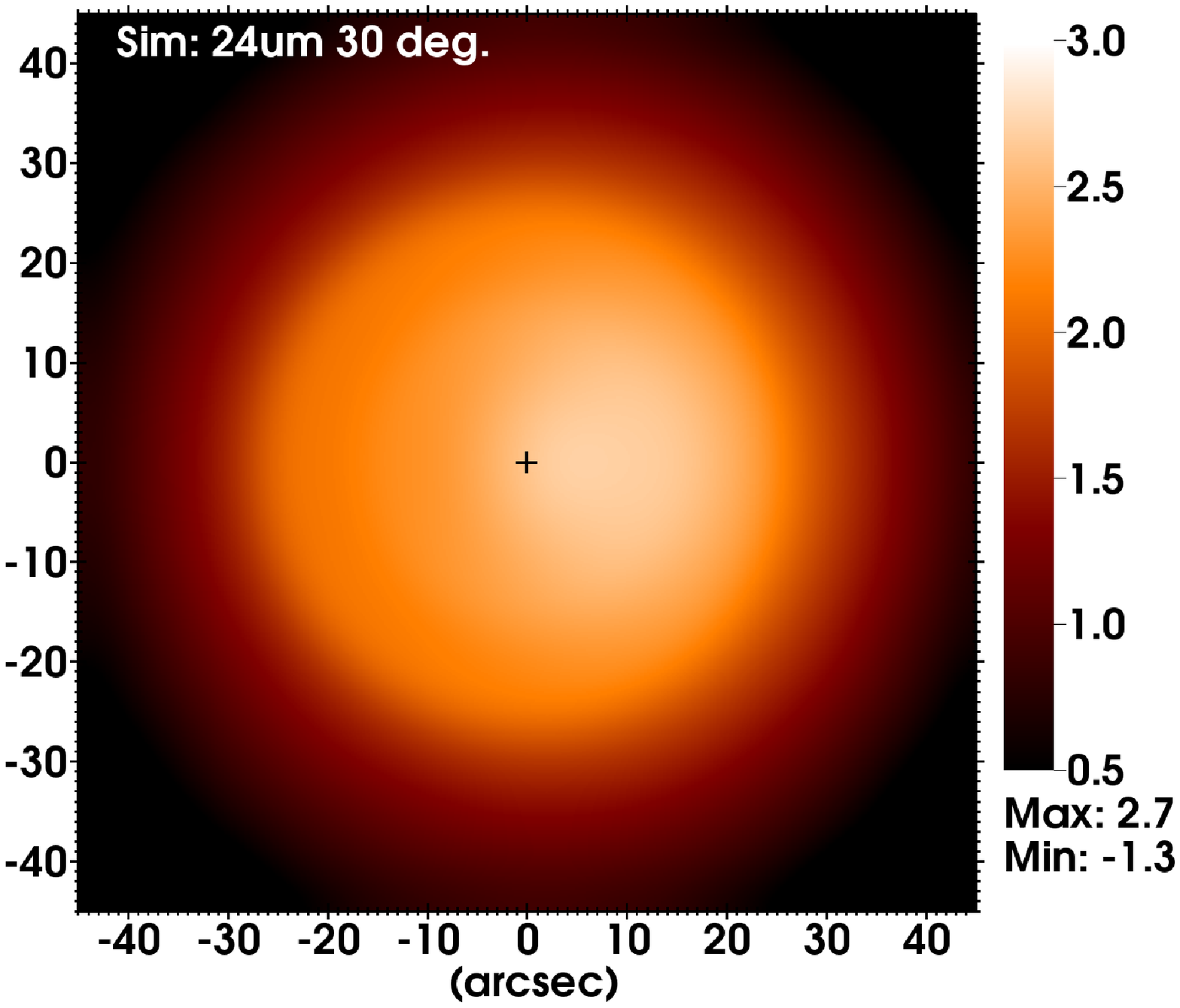}
\includegraphics[width=0.33\textwidth]{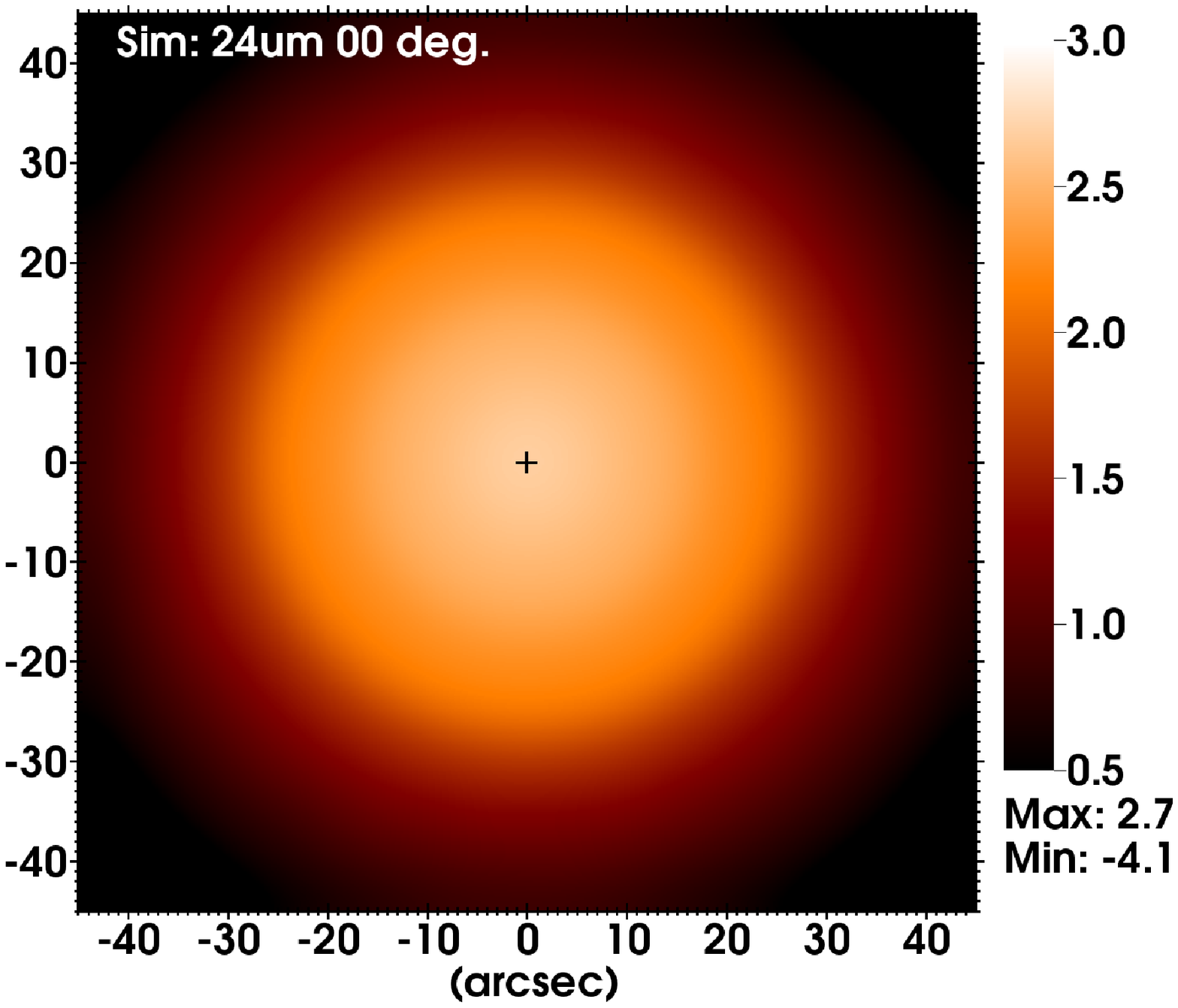}
\includegraphics[width=0.33\textwidth]{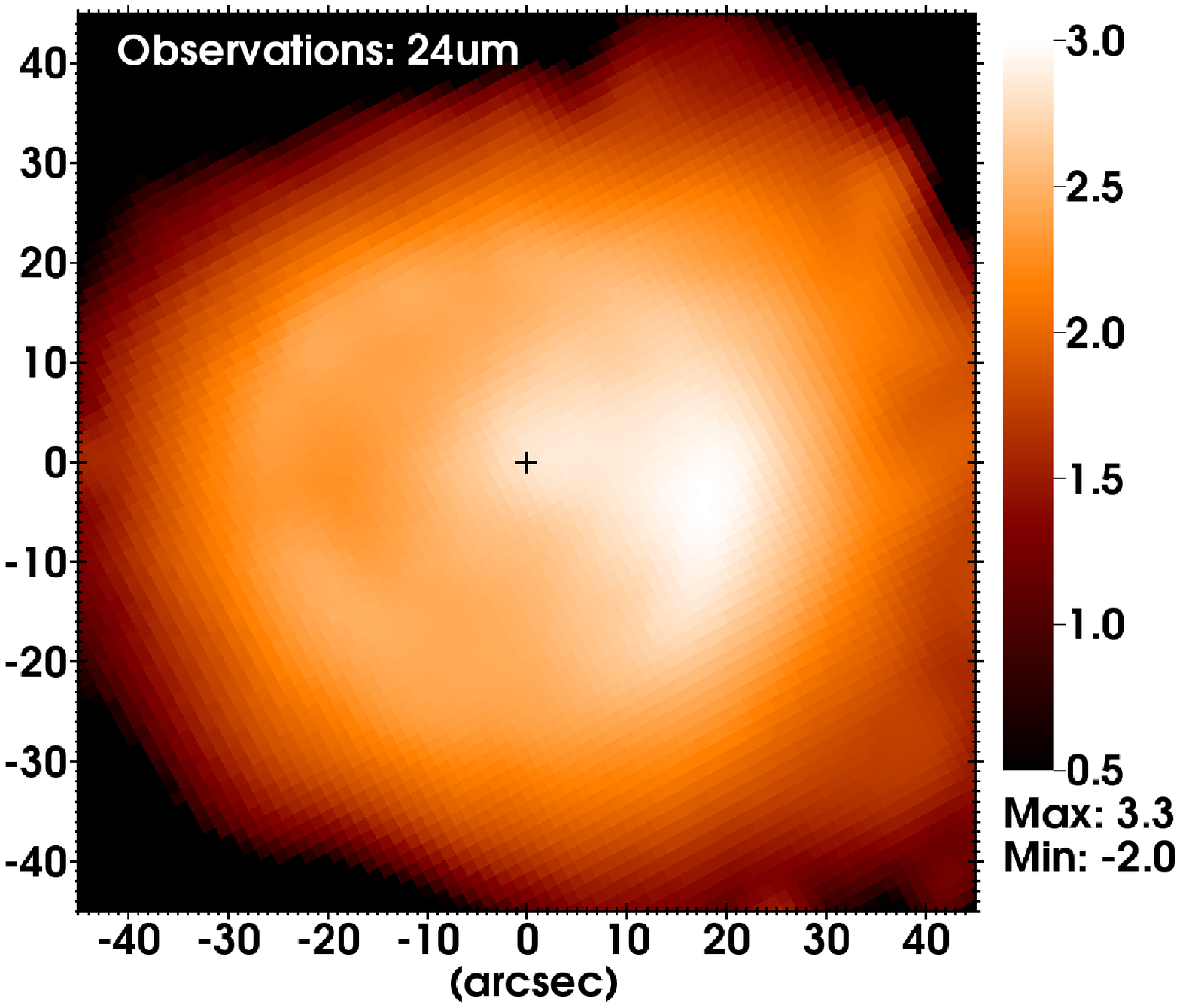}
\caption{Synthetic dust emission maps at 24\,$\mu$m of the simulated wind bubble and \hii region, 
calculated by {\sc torus} for Draine silicates. The panels show projections with a line-of-sight 
at angles of 90$\degr$, 60$\degr$, 45$\degr$, 30$\degr$, and 0$\degr$ to the symmetry axis of the 
simulation (left to right, top to bottom). The observational data are shown in the bottom right panel.
The colour bar shows intensity at 24\,$\mu$m on a log scale: $\log (I_{24\mu{\rm m}}/[{\rm MJy} \,
{\rm ster}^{-1}])$. Synthetic images are smoothed to an angular resolution of the MIPS instrument of 
6 arcsec (FWHM). A cross at the origin shows the location of the B1 star.}
\label{fig:simsIR}
\end{figure*}

Infrared emission from dust is calculated by postprocessing snapshots from the simulation 
using the {\sc torus} code (Harris 2000, 2015; Kurosawa et al. 2015) following the method 
described in Mackey et al. (2016). A very similar approach using {\sc torus} has also been 
used to model bow shocks around O stars (Acreman, Stevens \& Harries 2016). Briefly, we use 
{\sc torus} as a Monte-Carlo radiative transfer code to calculate the radiative equilibrium 
temperature of dust grains throughout the simulation, assuming radiative heating by the 
central star. The dust-to-gas ratio is assumed to be 0.01 in the ISM material (Draine et al. 2007), 
and we use a Mathis, Rumpl \& Nordsieck (1977) grain-size distribution defined by the minimum 
(maximum) grain size $a_{\rm min}=0.005\,\mu$m ($a_{\rm max}=0.25\,\mu$m) and a power law index 
$q=3.3$. The calculations here are for spherical silicate grains (Draine \& Lee 1984).
{\sc torus} calculates the dust temperature at all positions and then produces synthetic 
emission maps at different wavelengths from arbitrary viewing angles. Emission maps at 
24\,$\mu$m are shown in Fig.\,\ref{fig:simsIR} for 5 different viewing angles, and the observational 
image from {\it Spitzer} is shown on the same scale for comparison. The projection where the 
line-of-sight is at 60$\degr$ to the symmetry axis of the simulation provides a reasonable 
match to the observations, both in terms of morphology and absolute brightness. The exception 
is that the observational image seems to have emission from the position of the star itself, absent 
from our models. 

\begin{figure}
\center{\includegraphics[width=0.8\linewidth]{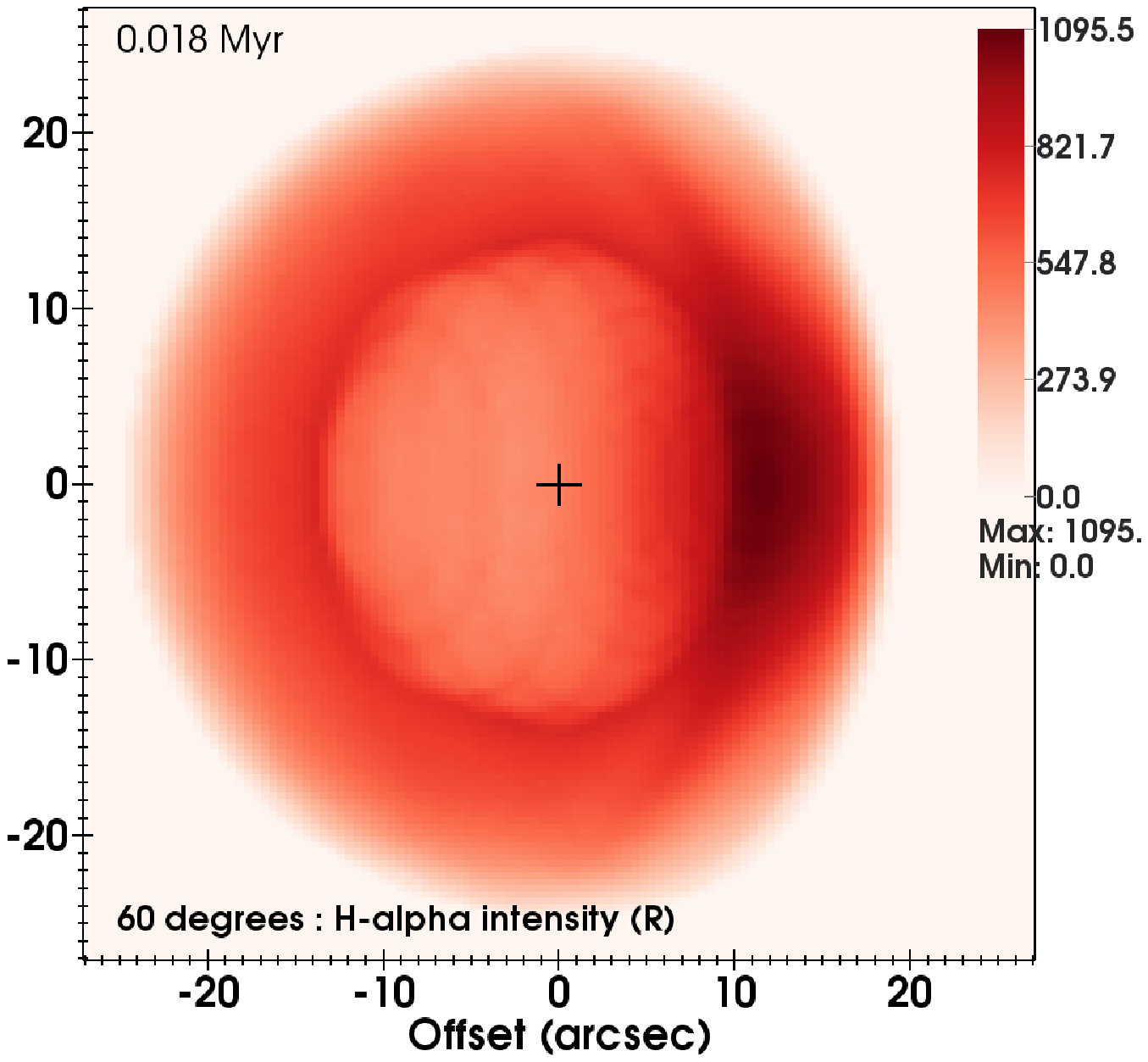}} {} \\
\center{\includegraphics[width=0.8\linewidth]{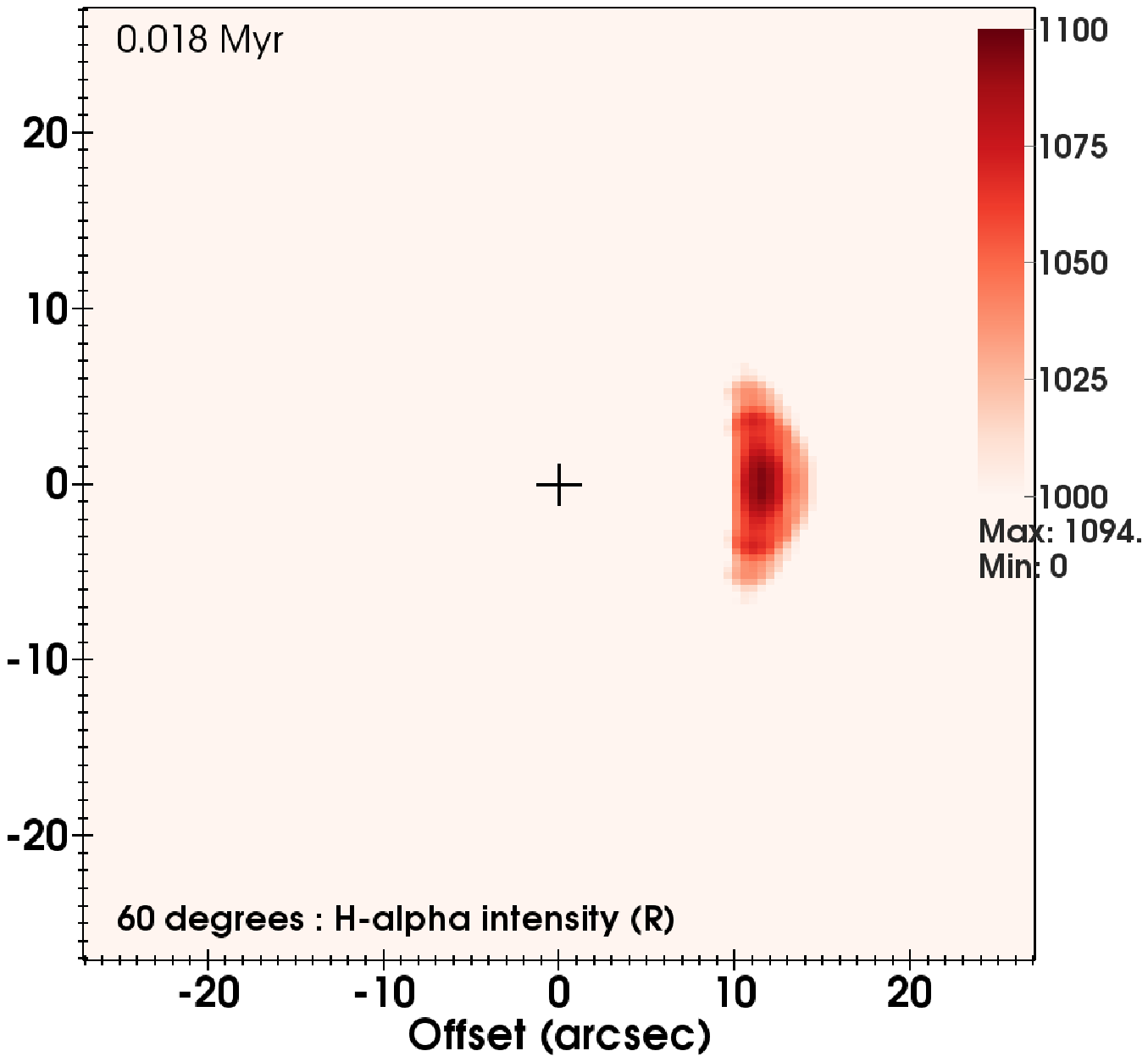}} {} \\
\vfill
\caption{H$\alpha$ emission maps calculated with {\sc visit}, for an angle of 60$\degr$\ to the symmetry axis 
of the simulation. The upper panel shows the predicted H$\alpha$ surface brightness in R on a linear scale
(although the absolute brightness should be treated with caution). The bottom panel shows only the brightest
10 per cent of the predicted emission, to mimic the case where the brightest emission is just above the 
noise level. A cross at the origin shows the location of the B1 star.} 
\label{fig:simsHA}
\end{figure}

We also calculated H$\alpha$ emission from the simulation at a projection angle of 60$\degr$, 
because this angle was the best match to the 24\,$\mu$m data. We did this using {\sc visit} (Childs et al. 
2012) and, while the relative brightness of each pixel is correctly calculated, there may be some scaling 
offset in the overall normalization of the image, so the absolute brightness should be treated with caution.
The result is plotted in Fig.\,\ref{fig:simsHA}, where the left-hand panel shows the emission map over the
full range of brightnesses and the right-hand one shows only the brightest 10 per cent of the predicted 
emission, to mimic the case where the brightest emission is just above the noise level (see 
Section\,\ref{sec:arc}). The brightest part of the simulated H$\alpha$ nebula is located within the
\hii region, about 10--15 arcsec from the star and well within the infrared bubble. This agrees very well 
with the observations in Fig.\,\ref{fig:neb}. Furthermore, Fig.\,\ref{fig:Ha} shows that the arc is about a
factor of 2 brighter than the rest of the \hii region in H$\alpha$ emission, comparable to what is shown 
in the left panel of Fig.\,\ref{fig:simsHA}.

Taken together, the agreement of the synthetic infrared and optical emission maps with the observational 
data is very encouraging. In particular, it suggests that H$\alpha$ imaging that is a factor of a few 
deeper would give a much clearer picture of the structure of this nebula, and strongly test our assertion 
that we are seeing a wind bubble and \hii region from a main-sequence B star. The simulation setup is very 
simple, with little fine-tuning to match the observations. If further observations support the interpretation 
as a young stellar wind bubble, then more detailed 3D simulations including a clumpy or turbulent medium and
a more realistic density structure of the molecular cloud would be clearly warranted. Particularly, one can 
expect that the presence of the denser material to the north-west of IRAS\,18153$-$1651 
(hub--N) would result in a stronger photoevaporation flow from this direction, which would explain the observed offset 
of stars\,1 and 2 from the geometric centre of the mid-infrared bubble as well as from that of the optical arc (cf. 
Ngoumou et al. 2013). 

\section{Summary and conclusion}
\label{sec:sum}

In this paper, we have reported the study of a compact, almost circular mid-infrared nebula associated with the IRAS 
source 18153$-$1651 and located in the region of ongoing massive star formation M17\,SWex. The study was motivated by the 
discovery of an optical arc near the centre of the nebula, leading to the hypothesis that it might represent a bubble
blown by the wind of a young massive star. To substantiate this hypothesis, we obtained optical spectra of the arc and 
two stars near its focus with the Gemini-South and the 3.5-m telescope in the Observatory of Calar Alto (Spain). The stars 
have been classified as main-sequence stars of spectral type B1 and B3, while the line ratios in the spectrum of the arc 
indicated that its emission is due to photoionization {\changed \bf $<...>$}. These findings allowed us to suggest 
that we deal with a recently formed low-mass ($\sim100 \, \msun$) star cluster (with the two B stars being its most massive 
members) and that the optical arc and IRAS\,18153$-$1651 are, respectively, the edge of a wind bubble and the \hii region 
produced by the B1 star. We also suggested that the one-sided appearance of the wind bubble is the result of interaction 
between the bubble and a photoevaporation flow from a molecular cloud located to the west of IRAS\,18153$-$1651 (the presence
of such a cloud is evidenced by submillimetre and radio observations). We supported our suggestions by simple analytical estimates, 
showing that the radii of the bubble and the \hii region would fit the observations if the age of IRAS\,18153$-$1651 is 
$\sim10\,000$\,yr and the number density of the ambient medium is $\sim100 \, {\rm cm}^{-3}$.
These estimates were validated by two-dimensional, radiation-hydrodynamics simulations investigating the effect of a 
nearby dense cloud on the appearance of newly forming wind bubble and \hii region. We found that synthetic 
H$\alpha$ and 24\,$\mu$m dust emission maps of our model wind bubble and \hii region show a good match to the observations,
both in terms of morphology and surface brightness. 

To conclude, taken together our results strongly suggest that we have revealed the first example of a wind bubble 
blown by a main-sequence B star.

\section{Acknowledgements} 
VVG acknowledges the Russian Science Foundation grant 14-12-01096. JM acknowledges 
funding from a Royal Society--Science Foundation Ireland University Research 
Fellowship (No. 14/RS-URF/3219). AYK acknowledges support from the National Research 
Foundation (NRF) of South Africa and the Russian Foundation for Basic Research grant 
16-02-00148. TJH is funded by an Imperial College Junior Research Fellowship. This 
research was supported by the Gemini Observatory, which is operated by the Association 
of Universities for Research in Astronomy, Inc., on behalf of the international Gemini 
partnership of Argentina, Brazil, Canada, Chile, and the United States of America, and 
has made use of the NASA/IPAC Infrared Science Archive, which is operated by the Jet 
Propulsion Laboratory, California Institute of Technology, under contract with the 
National Aeronautics and Space Administration, the SIMBAD data base and the VizieR 
catalogue access tool, both operated at CDS, Strasbourg, France.

\end{document}